\documentclass[10pt]{article}

\usepackage{amsmath}
\usepackage{amssymb}

\usepackage{graphicx}

\usepackage{cite}

\usepackage{color}

\usepackage{setspace}
\usepackage{subfigure}
\usepackage{url}

\newcommand{\sT}{S_{T}}
\newcommand{\sA}{S_{A}}
\newcommand{\dN}{D_{N}}
\newcommand{\dA}{D_{A}}

\usepackage{caption}

\makeatletter
\renewcommand{\@biblabel}[1]{\quad#1.}
\makeatother

\date{}

\begin{document}

\begin{flushleft}
{\Large
\textbf{Reaction-Diffusion-Delay Model for EPO/TNF-$\alpha$ Interaction in Articular Cartilage Lesion Abatement}
}
\\
Jason M.\ Graham$^{1,\ast}$,
Bruce P.\ Ayati$^{1}$,
Lei Ding$^{2}$,
Prem S.\ Ramakrishnan$^{2}$,
James A.\ Martin$^{2}$
\\
{\bf{1}} Department of Mathematics/Program in Applied Mathematical and Computational Sciences, University of Iowa, Iowa City, Iowa, USA
\\
{\bf{2}} Department of Orthopaedics and Rehabilitation, University of Iowa Hospitals and Clinics, Iowa City, Iowa, USA
\\
$\ast$ E-mail: jason-graham@uiowa.edu
\end{flushleft}

\section*{Abstract}

 Injuries to articular cartilage result in the development of lesions that form on the surface of the cartilage. Such lesions are associated with articular cartilage degeneration and osteoarthritis. The typical injury
 response often causes collateral damage, primarily an effect of inflammation, which results in the spread of lesions beyond the region where the
 initial injury occurs. We present a minimal mathematical model based on known mechanisms to investigate the spread and abatement of such lesions. In particular we represent the ``balancing act'' between pro-inflammatory and anti-inflammatory cytokines that is hypothesized to be a principal mechanism
 in the expansion properties of cartilage damage during the typical injury response. We present preliminary results of \emph{in vitro} studies that confirm the anti-inflammatory activities of the cytokine erythropoietin (EPO). We assume that the diffusion of cytokines determine the spatial
 behavior of injury response and lesion expansion so that a reaction diffusion system involving chemical species and chondrocyte cell state population densities is a natural way to represent cartilage injury response. We present computational results using the mathematical model showing that our representation is successful in capturing much of the interesting spatial behavior of injury associated lesion development and abatement in articular cartilage. Further, we discuss the use of this model to study the possibility of using EPO as a therapy for reducing the amount of inflammation induced collateral damage to cartilage during the typical injury response. The mathematical model presented herein suggests that not only are anti-inflammatory cytokines, such as EPO necessary to prevent chondrocytes signaled by pro-inflammatory cytokines from entering apoptosis, they may also influence how chondrocytes respond to signaling by pro-inflammatory cytokines.


\section*{Introduction}

  Articular cartilage is composed of cells known as chondrocytes. Mechanical stress and injury is known to kill (via necrosis) chondrocytes and results
  in the formation of lesions on the cartilage surface \cite{buckwalter1994,freedman2002,martin2,martin1}. The limited capacity of chondrocytes to self-repair, together with certain aspects
  of the typical injury response such as inflammation, can cause the spread of lesions and development of osteoarthritis. Recent research \cite{brines2008} suggests that inflammatory cytokines such as tumor necrosis factor $\alpha$ (TNF-$\alpha$) play a significant role
  in causing the spread of cartilage lesions. Anti-inflammatory cytokines such as erythropoietin (EPO) play an antagonistic role to TNF-$\alpha$,
  limiting the area over which a lesion can spread by counteracting some of the effects of inflammation \cite{brines2008}. Moreover, the authors in \cite{brines2008} describe the potential use of EPO as a therapy for cartilage injury and lesion abatement

  \begin{figure}[t]
  \centering
  \includegraphics[width=3.5in,height=3.5in]{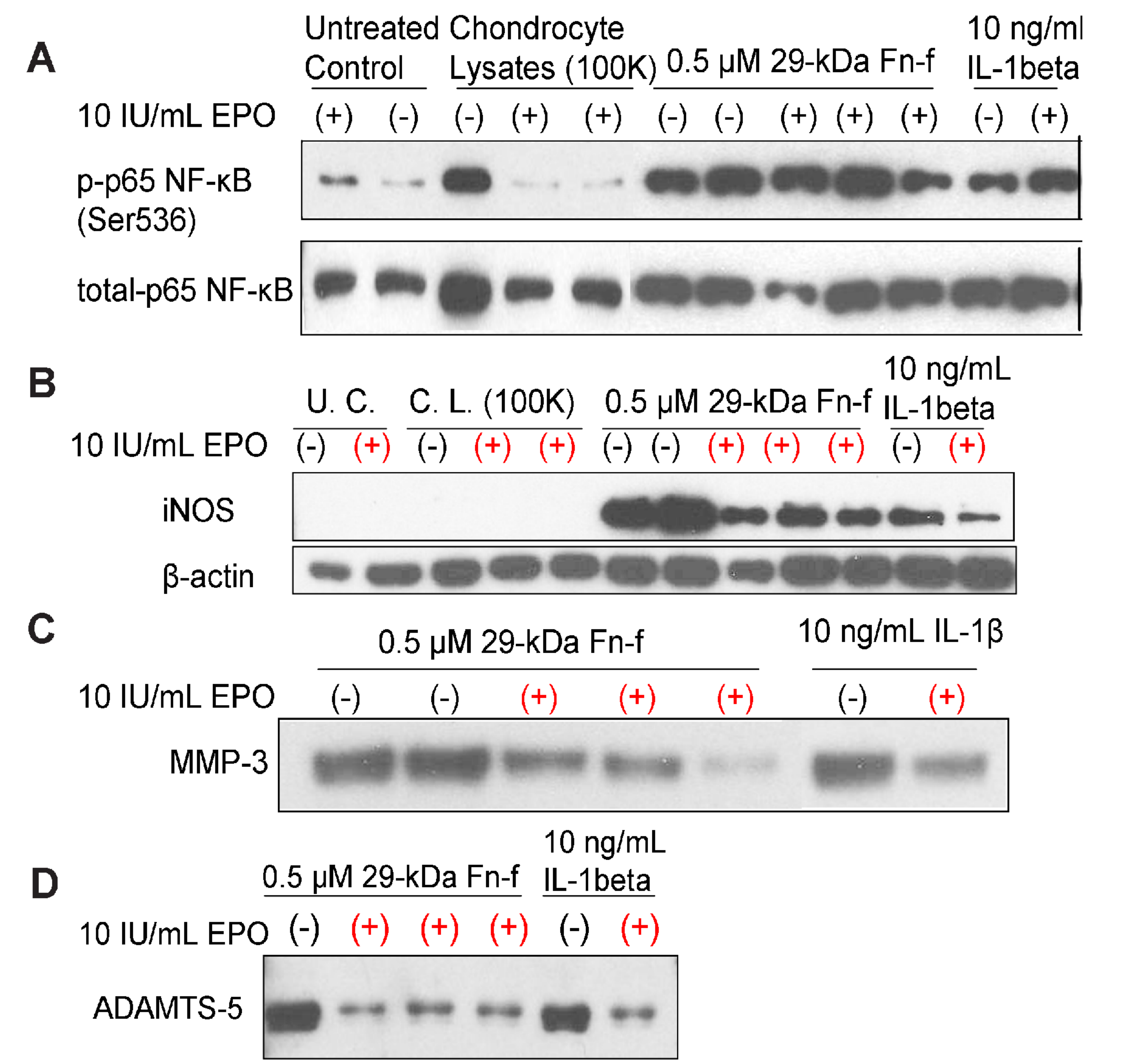}\\
  \caption{{\bf Effects of EPO on chondrocyte responses to alarmins and IL-1$\beta$. } Immunoblot
analyses are shown for phospho- and total nuclear factor kappa-B (NF-kB), a pro-inflammatory transcription factor (A), inducible nitric oxide synthase (iNOS), a catabolic signaling factor  (B), matrix metalloproteinase-3 (MMP-3) (C), and a disintegrin and metalloproteinase domain with thrombospondin repeats (ADAMTS-5) (D), which cause cartilage matrix degeneration. Primary cultures of bovine chondrocytes (1 x 106 cells/well) were treated with 10 IU/ml EPO for 24 hours, then challenged for an additional 24 hours with the indicated concentrations of purified 29 kDa fibronectin fragment, a known extracellular matrix DAMP, or interleukin-1b (IL-1b), a pro-inflammatory cytokine, or cell lysates containing multiple cellular DAMPS. EPO treatment suppressed catabolic responses induced by all three treatments. }\label{blots}
\end{figure}

  The potential for EPO to suppress catabolic and inflammatory responses of chondrocytes to various alarmins, i.e. molecules that trigger the innate immune response, was explored in a cell culture model (Figure \ref{blots}). Bovine articular chondrocytes in serum-free culture were pre-treated with EPO and then exposed to alarmins in the form of cell lysates or the 29 kDA fibronectin fragment (Fn-f). Interleukin-1$\beta$ (IL-1$\beta$)  was used as a control. Cell lysates were made by repeated freeze-thawing of excess cells from the same culture. Culture media were collected for immunoblot analysis of matrix metalloproteinase-3 (MMP-3) and A disintegrin and metalloproteinase with thrombospondin motifs 5 (ADAMTS-5). Cell layers were extracted for analysis of nuclear factor $\kappa$B (NF-$\kappa$B) and inducible nitric oxide synthases (iNOS). The results revealed that lysates, Fn-f, and IL-1$\beta$  induced NF-$\kappa$B phosphorylation (Figure \ref{blots}A). Fn-f, and IL-1$\beta$  but not lysates also induced iNOS, MMP-3 and ADAMTS-5 expression (Figure \ref{blots}B,C,D). In the case of NF- B, EPO treatment blocked activation by lysates, but not by Fn-f or IL-1 . EPO also dramatically reduced iNOS, MMP-3, and ADAMTS-5 expression in response to Fn-f and IL-1$\beta$. These results underscore the ability of EPO to counter injury effects on multiple pathways.

  \begin{figure}[t]
  \centering
  \includegraphics[width=2in,height=2in]{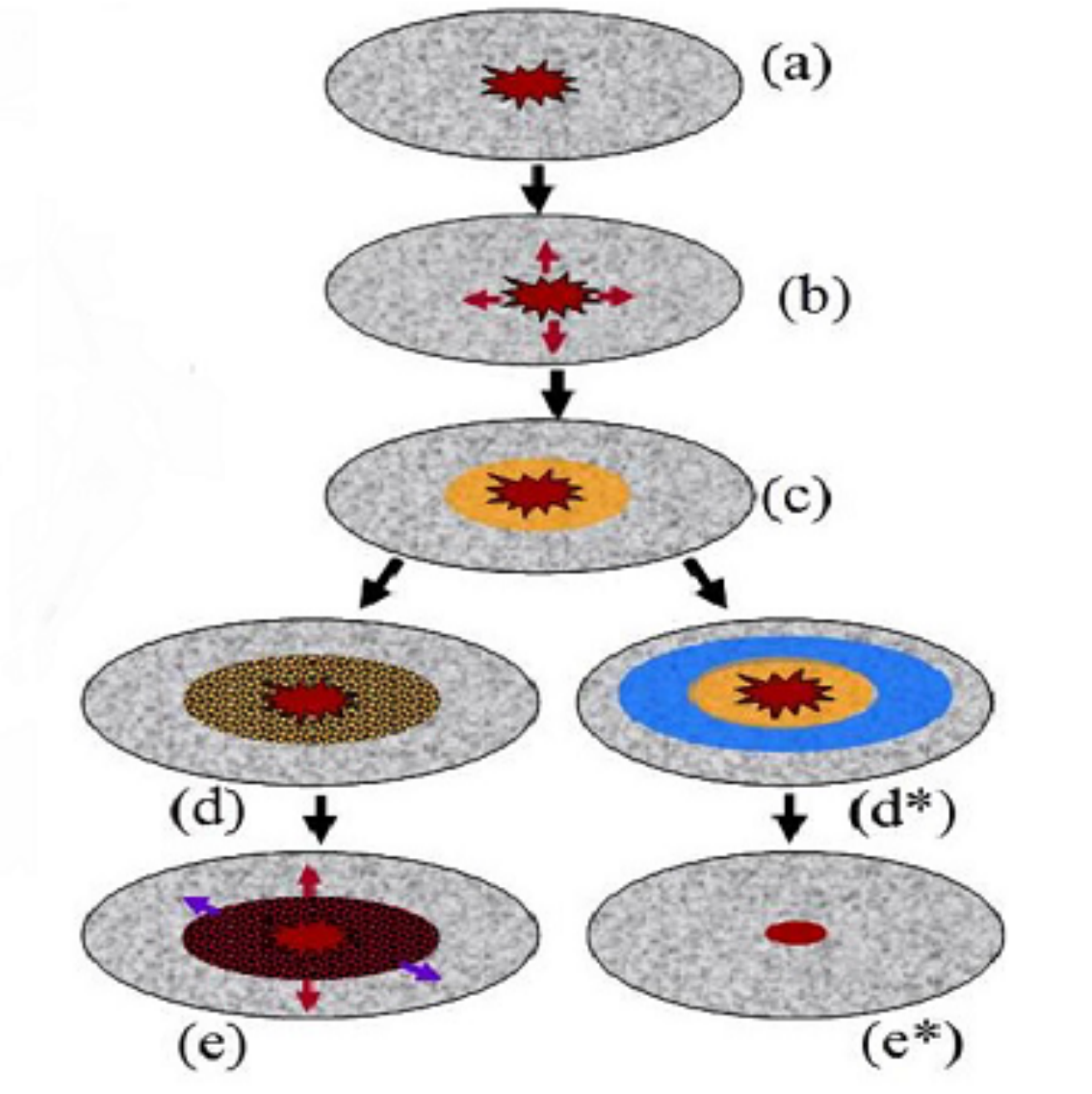}\\
  \caption{{\bf Injury response.} When a wound occurs DAMPs promote production of TNF-$\alpha$ (b), an initially viable penumbra (yellow in (c)) of cells becomes EPOR active. These cells are in danger of death which can result in the spread of injury as seen on the left in (d) and (e). However, EPO can counteract the effects of TNF-$\alpha$ and prevent spread of the lesion ($\text{d}^{\ast}$) and ($\text{e}^{\ast}$).}\label{epoTNF}
\end{figure}

 Figure \ref{epoTNF} illustrates the balance between pro-inflammatory (e.g. TNF-$\alpha$) and anti-inflammatory (e.g. EPO) cytokines during the typical injury response. Damage due to chemical or mechanical stress initiates production of alarmins, such as damage associated molecular pattern molecules (DAMPs) \cite{brines2008,harris2006,bianchi2006}. These alarmins trigger production of pro-inflammatory cytokines such as TNF-$\alpha$ by cells near the initial injury site (see (b) in figure \ref{epoTNF}) \cite{brines2008}. Thus there is a penumbra ((c) in figure \ref{epoTNF}) of ``sick'' cells formed at the boundary of the initial injury as a result of inflammation.  While still viable, these cells are at risk of dying themselves due to the presence of the pro-inflammatory cytokines, resulting in the spread of injury as shown in (d)-(e) in figure \ref{epoTNF}. However, cells far enough from the injury are able to produce anti-inflammatory cytokines such as EPO to check the spread of injury and promote healing. This is shown ($\text{d}^{\ast}$)-($\text{e}^{\ast}$) in figure \ref{epoTNF}. It is also suggested in \cite{brines2008} that TNF-$\alpha$ acts in such a way to limit EPO signaling. Thus there is a critical balance between pro-inflammatory and anti-inflammatory cytokines which determines the spreading properties of an articular cartilage lesion associated with injury. Moreover, the authors in \cite{brines2008} suggest that introduction of exogeneous EPO at an early time after an initial injury could be an effective therapy for minimizing the amount of collateral damage to cartilage that occurs as a result of inflammation.

The goal of this work is to study via mathematical modeling the EPO/TNF-$\alpha$ interaction described above. Herein we develop a model capable of simulating the scenario illustrated in figure \ref{epoTNF}. Moreover, the model is developed as a potential for studying the use EPO or other anti-inflammatory cytokines as a therapy for lesion abatement. The type of model, and in particular the one described below, in this article is of
value as it provides a link between observed phenomena and the mechanisms involved in driving that phenomena. Models such as those presented below
can also help to generate hypotheses and suggest experiments or fundamental quantities to be measured. More practically, mathematical modeling such as we use in this paper could provide quick and inexpensive initial screening of potential therapies.

\section*{Models}

 In this section we describe the mathematical model developed to represent the biological interactions of chondrocytes and cytokines described in \cite{brines2008}. We aim for a minimal model based on known mechanisms considered to be the dominant factors in articular cartilage lesion abatement. By a minimal model we mean one in which the removal of any component results in behavior that is inconsistent with the typical injury responses in cartilage as discussed in \cite{brines2008}. We note that the model described below can be expanded or modified to include further interactions as new results from experiment and observation dictate. In particular, further chemical pathways such as interleukin-1 $\beta$  IL1-$\beta$ can easily be incorporated \emph{if necessary}. The chemical species included below are chosen for their functionality, that is their action on chondrocytes during injury response. Hence they can be replaced with any other chemical species whose effects on chondrocytes are functionally the same. The chondrocyte cell states described below represent the biological actions of the cells in response to signaling by a particular cytokine during the typical injury response. These actions are assumed to be analogous to those of other cell types, for which aspects of innate immune response, such as local inflammation considered in this paper, are well established.

\begin{figure}[t]
  \centering
  \includegraphics[width=3in,height=1.5in]{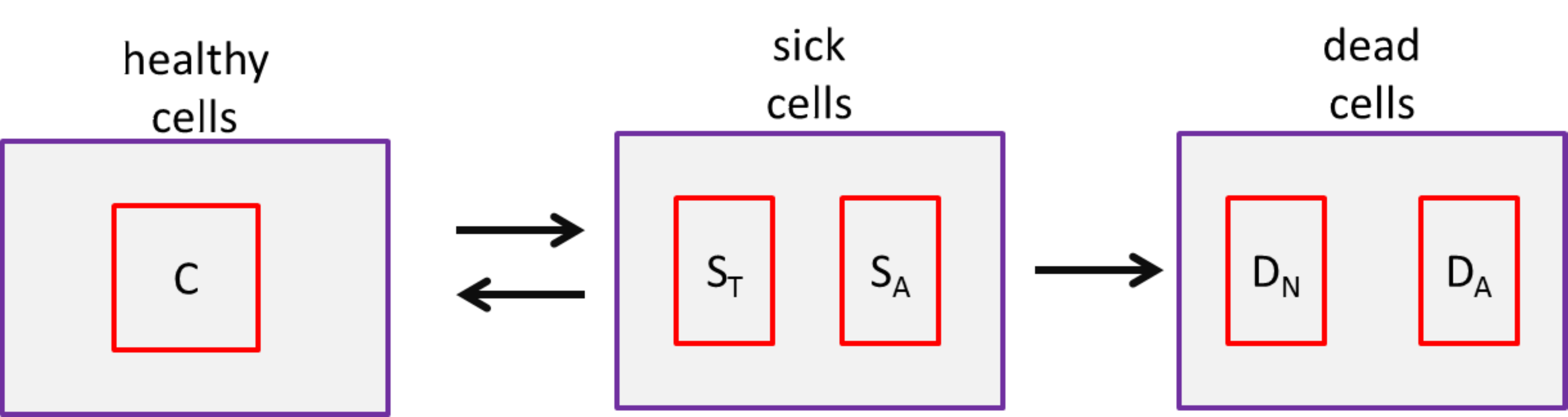}\\
  \caption{{\bf States of chondrocytes during stereotypical injury response.} Healthy cells ($C$) become sick. Typically a healthy cell will first become catabolic ($\sT$) and either enter apoptosis ($\dA$), or become EPOR active ($\sA$). EPOR active cells may be saved by EPO to become healthy once again, or die (apoptosis).}\label{states}
 \end{figure}

 We consider a population of chondrocytes fixed in matrix, which is typical of articular cartilage. Subpopulations of chondrocytes are assumed to exist in
 different states corresponding to the chemical signals being received by the cells during injury response. Figure \ref{states} shows the states in which subpopulations of chondrocytes may exist. We refer to the normal state of a subpopulation of chondrocytes as the healthy state. We denote by $C$ the population density (cells per unit area) of healthy chondrocytes at a given time and location. As a result of inflammation and injury healthy chondrocytes can enter into a ``sick'' state. Cells in this state are at risk death (via apoptosis) unless their signaling by TNF-$\alpha$ is somehow limited. We consider two subpopulations of cells in the sick state. We denote by $\sT$ the population density of cells in the ``catabolic'' state. Catabolic cells are chondrocytes that have been signaled by alarmins and are capable of synthesizing TNF-$\alpha$ and other cytokines associated with inflammation. Healthy cells signaled by DAMPs or TNF-$\alpha$ enter into the catabolic state and begin to synthesize TNF-$\alpha$ and produce reactive oxygen species (ROS). Catabolic cells that are signaled by TNF-$\alpha$ express a receptor (EPOR) for EPO and make up the subpopulation of sick cells we refer to as EPOR active. It should be noted that there is a time delay of 8--12 hours before a cell expresses the EPO receptor after being signaled to become EPOR active \cite{brines2008}. We denote the
 population density of EPOR active cells by $\sA$. Since EPOR active cells express a receptor for EPO, they may switch back to the healthy state if signaled by EPO. However, as discussed in \cite{brines2008} TNF-$\alpha$ limits production of EPO. Thus there is a balance between EPO and TNF-$\alpha$ that determines the spreading behavior of cartilage lesions. The catabolic and EPOR active cells together make up the population of cells forming the penumbra as illustrated in figure \ref{epoTNF}. We also consider a ``dead'' state for subpopulations of chondrocytes. This includes necrotic cells $\dN$ and apoptotic cells $\dA$. We note that for the purposes considered herein that apoptotic cells do not feed back into the system and thus are not explicitly represented in the mathematical model. We assume chondrocytes become necrotic only at the time of an initial injury. Necrotic cells release alarmins such as DAMPs that initiate the injury response. Either catabolic or EPOR active cells become apoptotic if signaled by inflammatory cytokines.

 \begin{figure}[t]
  \centering
  \includegraphics[width=2.8in,height=2.8in]{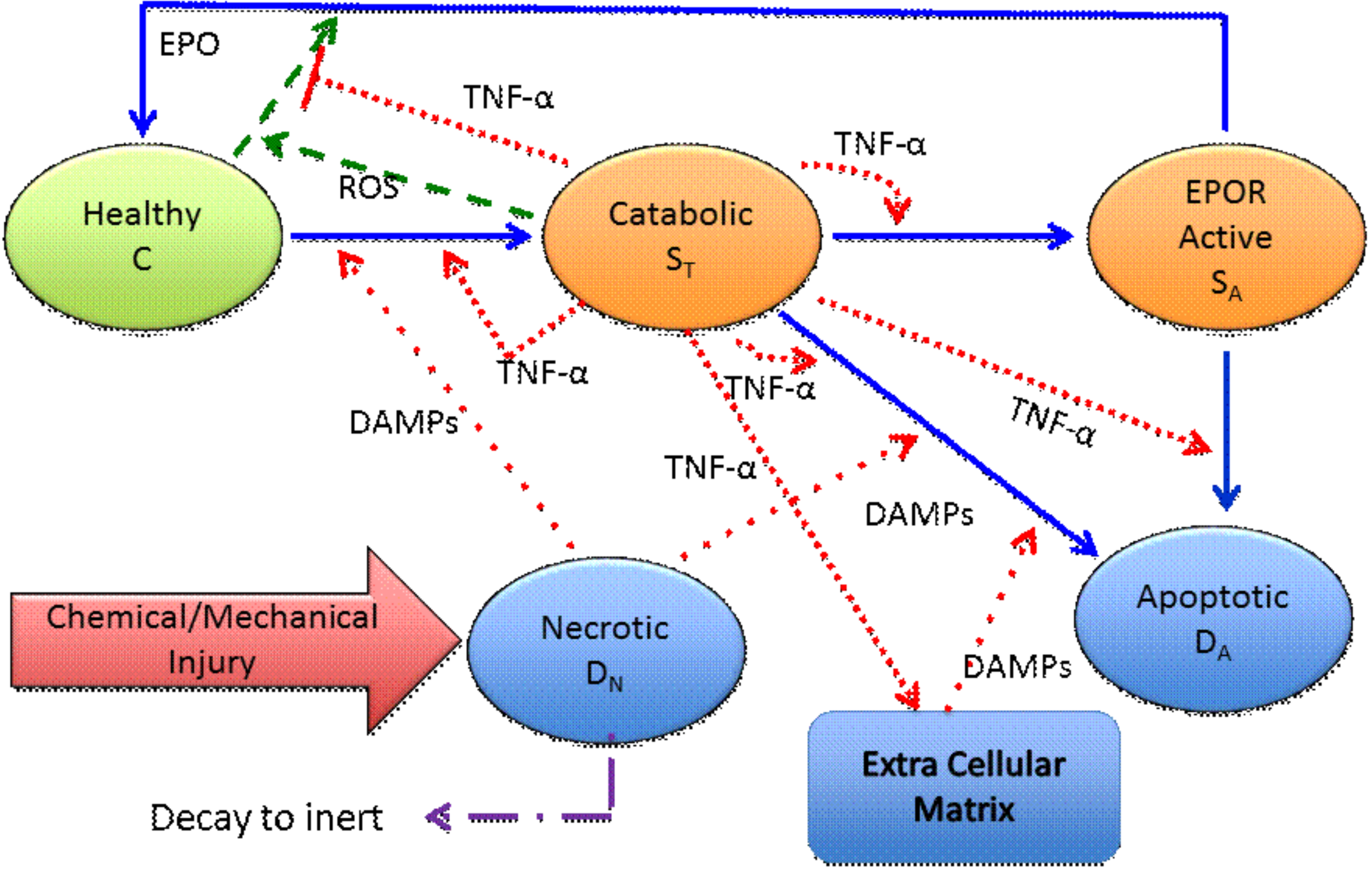}\\
  \caption{{\bf Signaling involved in cartilage injury response.} The occurrence of an injury begins a sequence of chemical productions that promote the inflammatory response. Lysing (necrotic) cells give off DAMPs ($M$) resulting in a population of catabolic cells that produce TNF-$\alpha$ ($F$) and ROS ($R$). ROS promotes production of EPO ($P$) by healthy cells while TNF-$\alpha$ acts to block EPO signaling. Furthermore, TNF-$\alpha$ degrades extracellular matrix providing another source of DAMPs. The cells switch state according to the signaling by chemicals produced in response to injury and inflammation. DAMPs and TNF-$\alpha$ drive the $C \rightarrow \sT$ transition, TNF-$\alpha$ drives the $\sT \rightarrow \sA$ transition, while TNF-$\alpha$ and DAMPs drive the $\sT \rightarrow \dA$ transition. Finally, EPO drives the $\sA \rightarrow C$ transition. This figure illustrates the assumptions of the mathematical model ((\ref{eq:fullSys}) below). Not shown: DAMPs given off as a result of matrix degradation also influences the switch from healthy to catabolic.}\label{signaling}
\end{figure}

  Figure \ref{signaling} details the chemical signaling and the switching of chondrocyte cell states represented in the mathematical model presented in the following. An initial injury creates a population of necrotic cells which release alarmins (such as DAMPs) \cite{harris2006,bianchi2006}. We denote by $M$ the concentration of DAMPs at a given time and location. The DAMPs signal healthy cells near the injury to enter the catabolic state resulting
  in the production of TNF-$\alpha$. We denote the concentration of TNF-$\alpha$ at a given time and location by $F$. The inflammatory cytokine TNF-$\alpha$ has several effects on the system: It
  \begin{enumerate}
    \item feeds back to continue to switch healthy cells into the catabolic state,
    \item causes catabolic cells to enter the EPOR active state \cite{brines2008},
    \item influences apoptosis of catabolic and EPOR active cells,
    \item degrades extracellular matrix (denoted by $U$) which results in increased concentrations of DAMPs,
    \item has a limiting effect on production of EPO \cite{brines2008}.
  \end{enumerate}
  Catabolic cells also produce reactive oxygen species (ROS) which influences the production of EPO by healthy cells. We denote the concentration of ROS at a given time and location by $R$. There is a time delay of 20--24 hours before a healthy cell signaled by ROS will begin to produce EPO.

 In developing a mathematical model to represent the scenario described above we assume that the chemicals diffuse throughout a domain. This diffusion is essential in determining the spatial behavior, i.e. the expansion or abatement of the lesion. Since the chondrocytes are fixed in the matrix we do not consider cell motility. First the mathematical model contains four equations (one for each chemical species) describing the dynamics of the chemical concentrations. These equations are each of the form:
 \begin{align}
    & \mbox{change in concentration of particular chemical} \nonumber \\
     & =  \mbox{diffusion of that chemical} \nonumber \\
    & + \mbox{production of that chemical by cells in appropriate states}\nonumber \\
    & - \mbox{natural decay of that chemical}.
 \end{align}
 Next the model consists of four equations for the population densities of cells in the healthy, catabolic, EPOR active, and necrotic states.
 Each of these equations are of the form:
 \begin{align}
   &\mbox{change in population density of particular cell state}  \nonumber \\
    & = \mbox{population density of cells switching into that state due to appropriate signaling} \nonumber \\
     & - \mbox{population density of cells switching out of that state due to appropriate signaling}.
 \end{align}
 Finally there is an equation corresponding to the degradation of extracellular matrix by TNF-$\alpha$ which feeds back into the production
 of DAMPs.

 The model equations for the concentrations of the chemical species are:
 \begin{subequations}
   \begin{align}
    \partial_{t}R = & \nabla\cdot(D_{R}\nabla R) - \delta_{R}R  + \sigma_{R}\sT, \label{eq:ros}\\
    \partial_{t}M = & \nabla\cdot(D_{M}\nabla M) - \delta_{M}M + \sigma_{M}\dN + \delta_{U}U\frac{F}{\lambda_{F} + F},\label{eq:damps}\\
    \partial_{t}F  = & \nabla\cdot(D_{F}\nabla F) - \delta_{F}F + \sigma_{F}\sT, \label{eq:tnf}\\
    \partial_{t}P  = &  \nabla\cdot(D_{P}\nabla P) - \delta_{P}P
    + \sigma_{P}C(t-\tau_{2})\frac{R(t-\tau_{2})}{\lambda_{R} + R(t-\tau_{2})}\frac{\Lambda}{\Lambda + F}\label{eq:epo}.
  \end{align}

  The equation describing matrix degradation by TNF-$\alpha$ is:
   \begin{align}
    \partial_{t}U = & -\delta_{U}U\frac{F}{\lambda_{F} + F}. \label{eq:matrix}
  \end{align}
  \label{eq:fullSys}
  We note that the right hand side of this equation appears in (\ref{eq:damps}) as part of the production of DAMPs as we have assumed that degraded matrix releases alarmins.

  The equations for the switching of cell states are:
   \begin{align}
    \partial_{t} C = & \alpha\sA\frac{P}{\lambda_{P} + P} - \beta_{1} C\frac{M}{\lambda_{M} + M}H(P-P_{c})-\beta_{2}C\frac{F}{\lambda_{F} + F}H(P-P_{c}),  \label{eq:healthy}\\
    \partial_{t}\sT = & \beta_{1} C\frac{M}{\lambda_{M} + M}H(P-P_{c})+\beta_{2}C\frac{F}{\lambda_{F} + F}H(P-P_{c})- \gamma\sT(t-\tau_{1})\frac{F(t-\tau_{1})}{\lambda_{F} + F(t-\tau_{1})} \nonumber \\
     &-\nu\sT\frac{F}{\lambda_{F} + F}\frac{M}{\lambda_{M} + M} , \label{eq:cata}\\
  \partial_{t}\sA = & \gamma\sT(t-\tau_{1})\frac{F(t-\tau_{1})}{\lambda_{F} + F(t-\tau_{1})} - \alpha\sA\frac{P}{\lambda_{P} + P} - \mu_{\sA}\sA\frac{F}{\lambda_{F} + F} , \label{eq:epor}\\
  \partial_{t}\dN = &  - \mu_{\dN}\dN. \label{eq:necrot}
  \end{align}
  \label{eq:fullSys}
  \label{eq:fullSys}
  \end{subequations}
  In (\ref{eq:healthy}) and (\ref{eq:cata}) the function $H(\cdot)$ is given by
  \begin{align}
   H(s) = & \left\{ \begin{array}{ll} 1 & \mbox{if $s < 0$,} \\ 0 & \mbox{if $s \geq 0$.}\end{array} \right. \label{eq:Hswitch}
  \end{align}
  The constant $P_{c}$ represents a critical level of EPO above which the effects of DAMPs and TNF-$\alpha$ on healthy cells is limited.

  Based on the diagram in figure \ref{signaling} and the assumptions underlying that diagram, one could assume that the function $H(\cdot)$ is identically one, i.e. $H \equiv 1$. However, if we take $H \equiv 1$ then figure \ref{signaling} implies, and computational results confirm, that
  there is infinite feedback into the system by alarmins, TNF-$\alpha$, and catabolic cells. As noted in \cite{brines2008} ``the pro-inflammatory arm of the injury response is inherently self-amplifying''. This does not allow for lesion abatement
  or limitation of secondary pro-inflammatory cytokine induced injury described in \cite{brines2008}. This suggests that signaling by anti-inflammatory cytokines such as EPO not only promote the switch from the EPOR active state to the healthy state but also influences the response of healthy cells to alarmins and pro-inflammatory cytokines. Thus the mathematical model suggests that in some way the anti-inflammatory cytokines limit the switch from the healthy state to the catabolic state so that there is not an infinite feedback into the system by alarmins, TNF-$\alpha$, and catabolic cells. This is consistent with the observation of Brines and Cerami on the antagonistic relationship between TNF-$\alpha$ and EPO, that ``each is capable of suppressing the biological activity of the other''\cite{brines2008}. In the model equations (\ref{eq:fullSys}) we take the function $H(\cdot)$  as in (\ref{eq:Hswitch}) for convenience. However, there may be a more appropriate form for the function $H(\cdot)$ that must be determined through experiment to discover the nature of the effects that sufficient concentrations of EPO or other anti-inflammatory cytokines have on healthy chondrocytes.


  We note some features of the mathematical model {\ref{eq:fullSys}}:
  \begin{enumerate}
    \item we have incorporated the time delays for activation of the EPO receptor and synthesis of EPO (\ref{eq:epo}),(\ref{eq:cata}),(\ref{eq:epor}),
    \item it requires a concentration of TNF-$\alpha$ and DAMPs together for apoptosis of catabolic cells (see (\ref{eq:cata})),
    \item after some time necrotic cells decay to an inert state (see (\ref{eq:necrot})) so that there is not a continuous
    production of DAMPs for all time from the initial injury. This also corresponds to the loss of cartilage such as is sometimes associated with osteoarthritis.
  \end{enumerate}
  The specific functional forms appearing in the model system (\ref{eq:fullSys}) have been chosen to capture the critical thresholds and represent the function shapes qualitatively.

\section*{Results and Discussion}

  Using the mathematical model (\ref{eq:fullSys}) we can simulate the two scenarios depicted in figure \ref{epoTNF}. That is, we can choose parameter
  values that correspond first to the case where inflammation is not limited by the presence of anti-inflammatory cytokines  (d) and (e) in figure \ref{epoTNF} and second to the case where anti-inflammatory cytokines limit the development of secondary inflammation induced injury  ($\text{d}^{\ast}$) and ($\text{e}^{\ast}$) in figure \ref{epoTNF}. For the computations and simulations we assume a two dimensional dimensional domain. Moreover, as depicted in figure \ref{epoTNF} we assume circular symmetry so that changes only occur in the radial direction. In order to simulate the injury response we obtain numerical approximations to the system of partial differential equations (\ref{eq:fullSys}). This is done
  by discretizing the system in the spatial terms using a standard finite difference scheme appropriate for circular symmetry. This results in a system of delay-differential equations which are solved in MATLAB using the {\tt dde23} code for solving delay-differential equations. For details on the methods and software for solving delay-differential equations see \cite{larry1,larry2,larry3}.

  First, using the mathematical model (\ref{eq:fullSys}) we simulate the evolution of chondrocyte population density (per area, and assuming circular symmetry, as a function of radius) over a period of ten days after an initial cartilage injury. This corresponds to the cascade of events illustrated in the left hand arrows of figure \ref{epoTNF}. The initial conditions correspond to the presence of an initial injury, i.e. a population of necrotic cells occupying a circle of radius 0.25 mm.

\begin{figure}
   \centering
   \subfigure[t=0]{\includegraphics[height=1.7in]{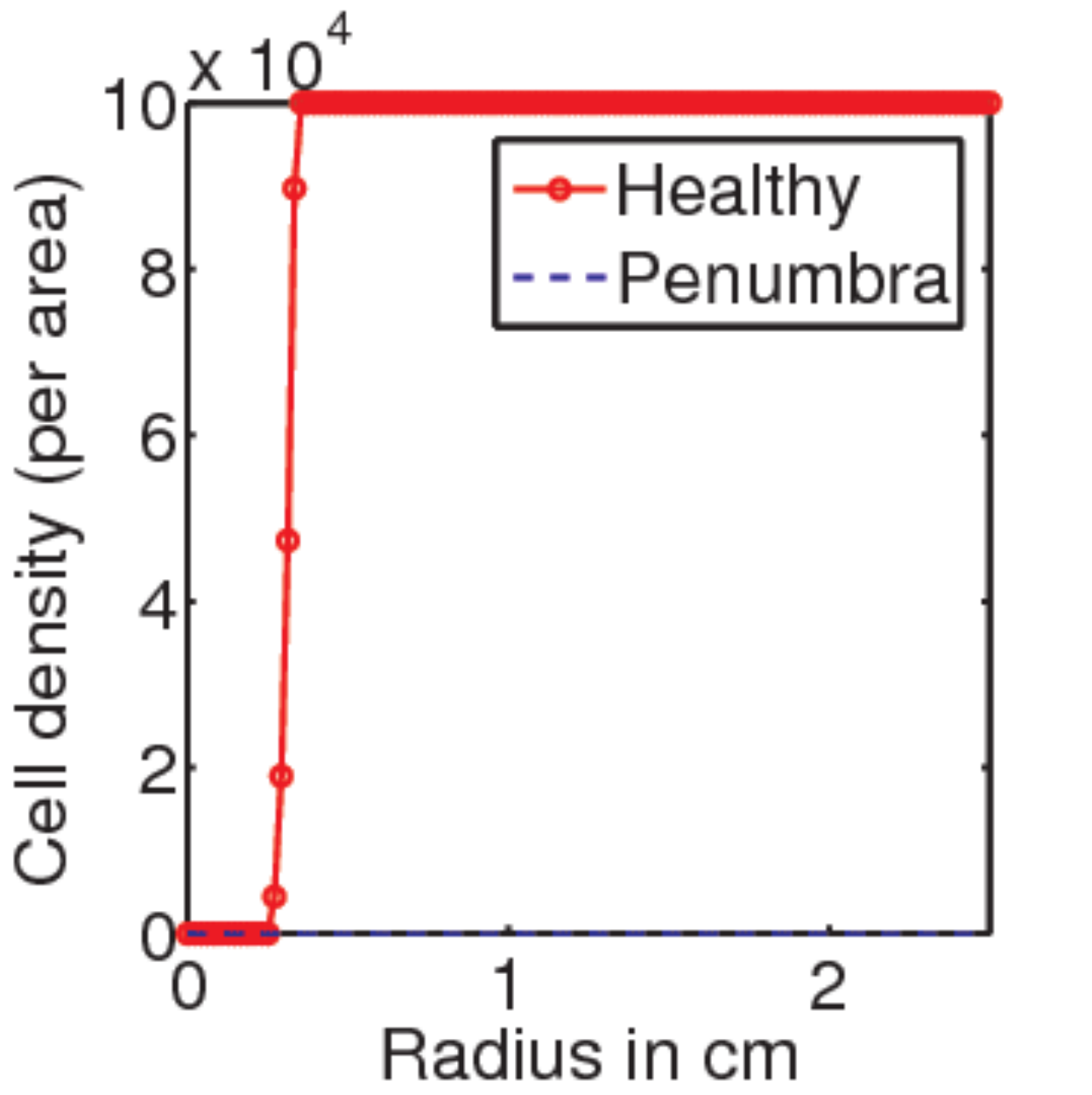}\label{den-a}}
   \subfigure[t=1.25 days]{\includegraphics[height=1.7in]{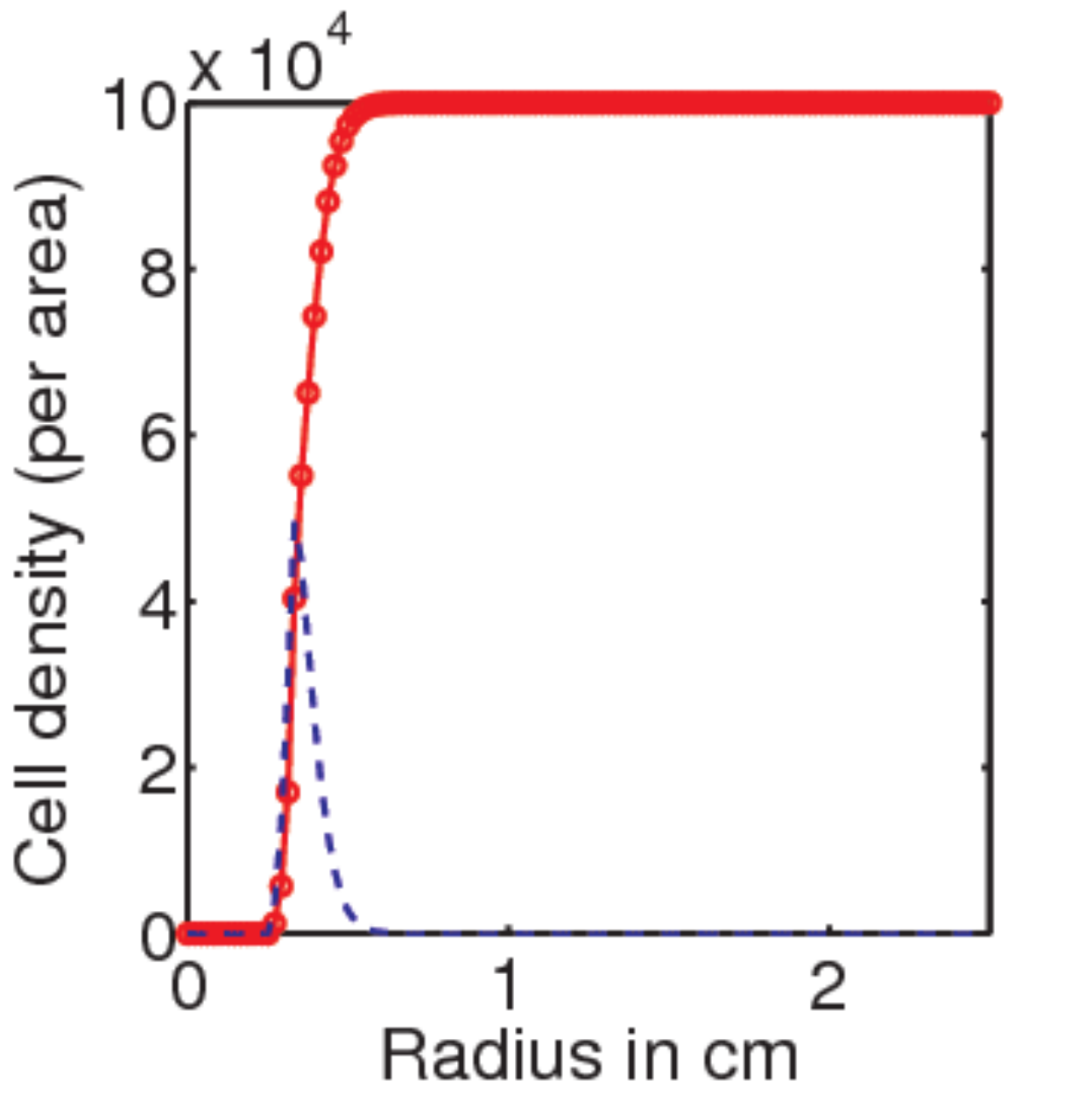}\label{den-b}}
   \subfigure[t=2.5 days]{\includegraphics[height=1.7in]{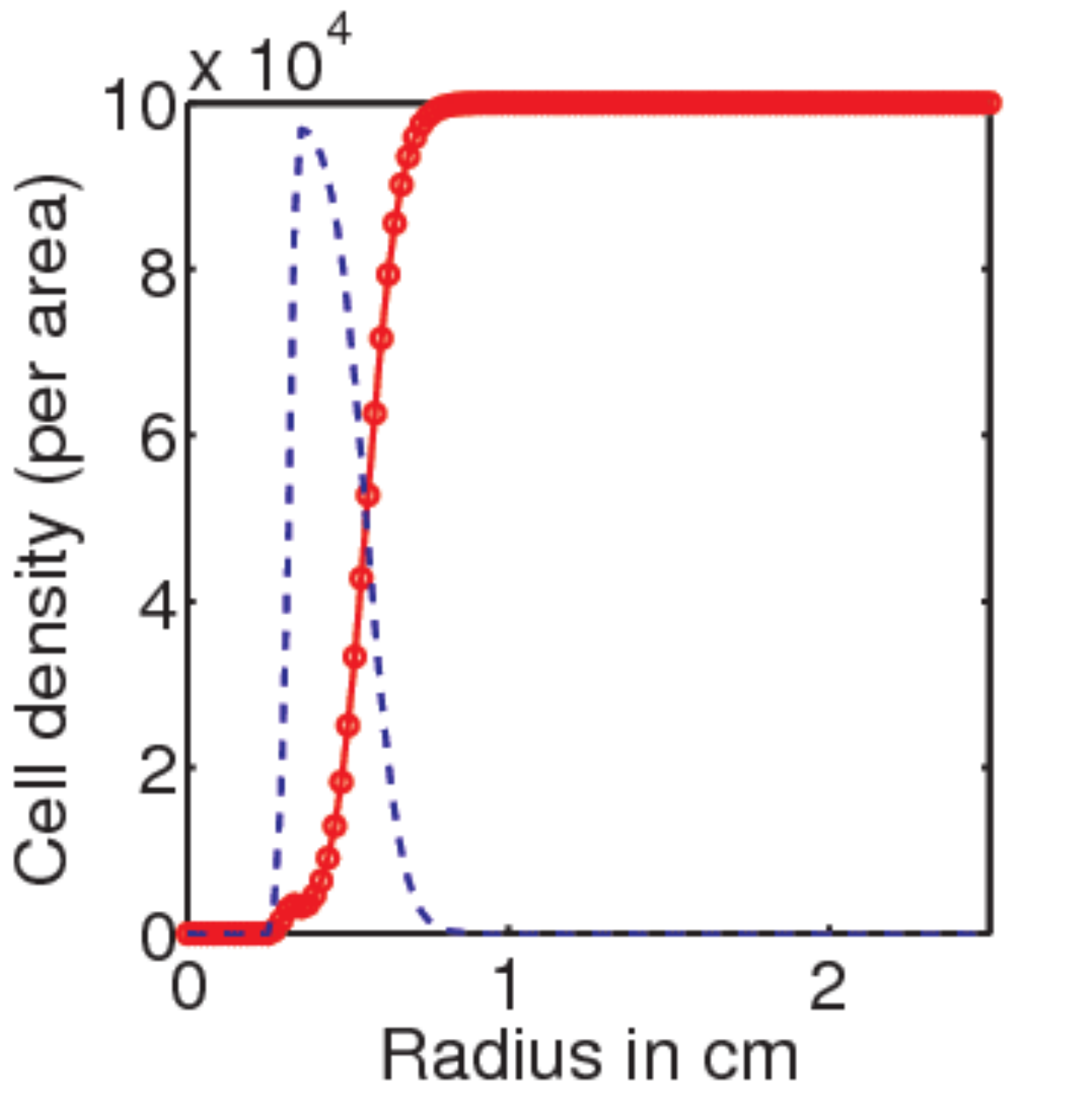}\label{den-c}}\\
   \subfigure[t=3.75 days]{\includegraphics[height=1.7in]{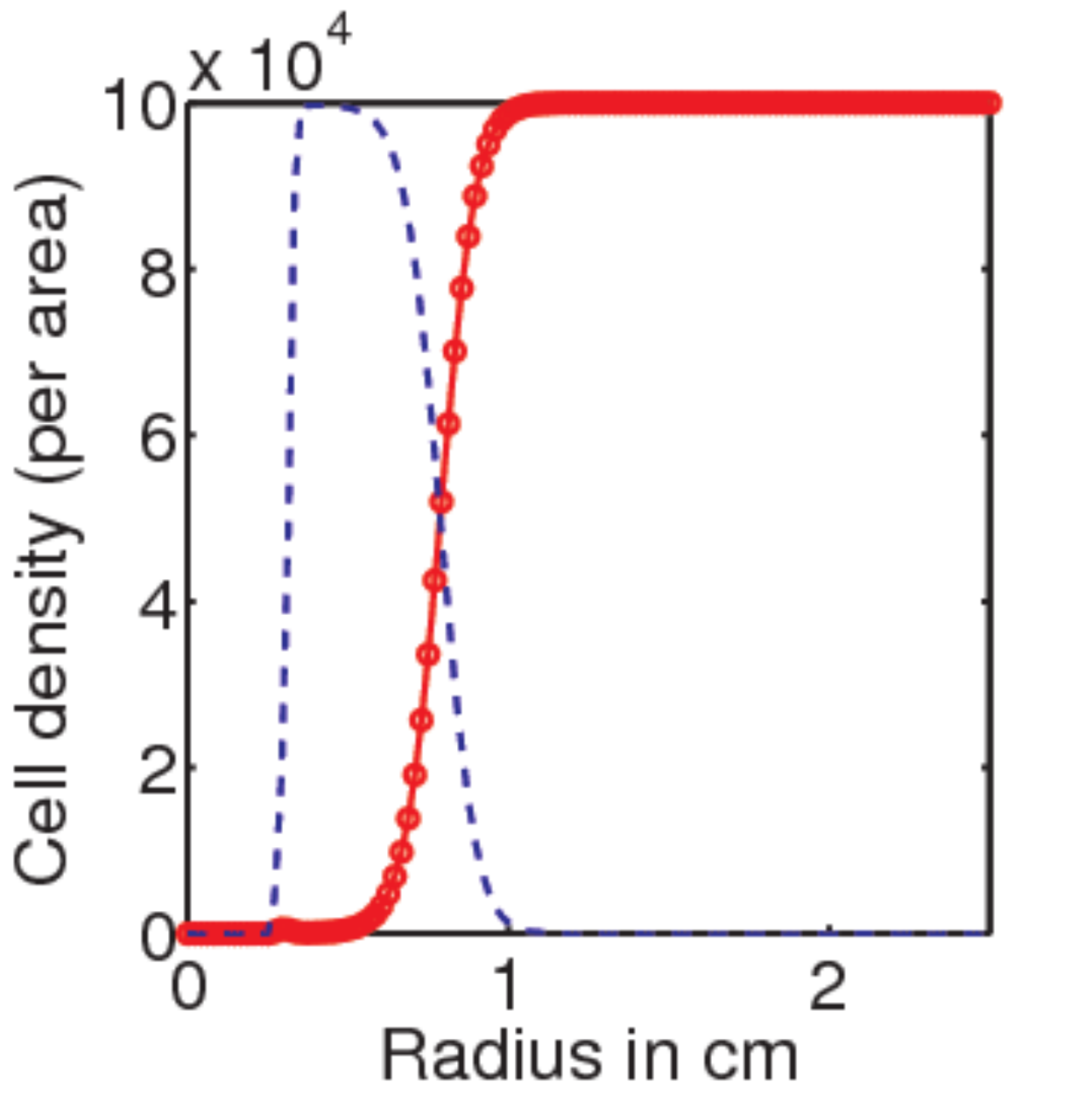}\label{den-d}}
   \subfigure[t=5 days]{\includegraphics[height=1.7in]{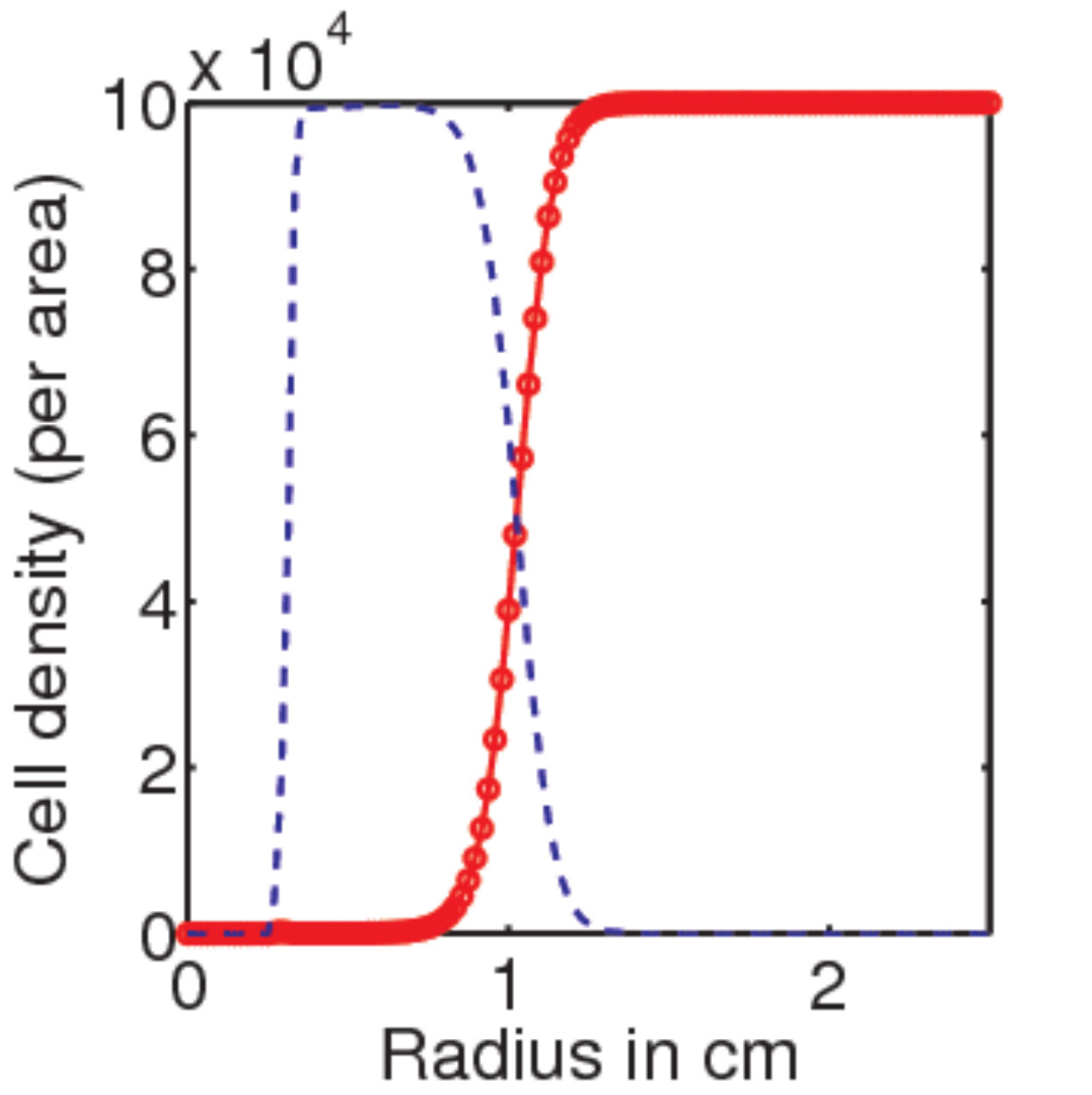}\label{den-e}}
   \subfigure[t=6.25 days]{\includegraphics[height=1.7in]{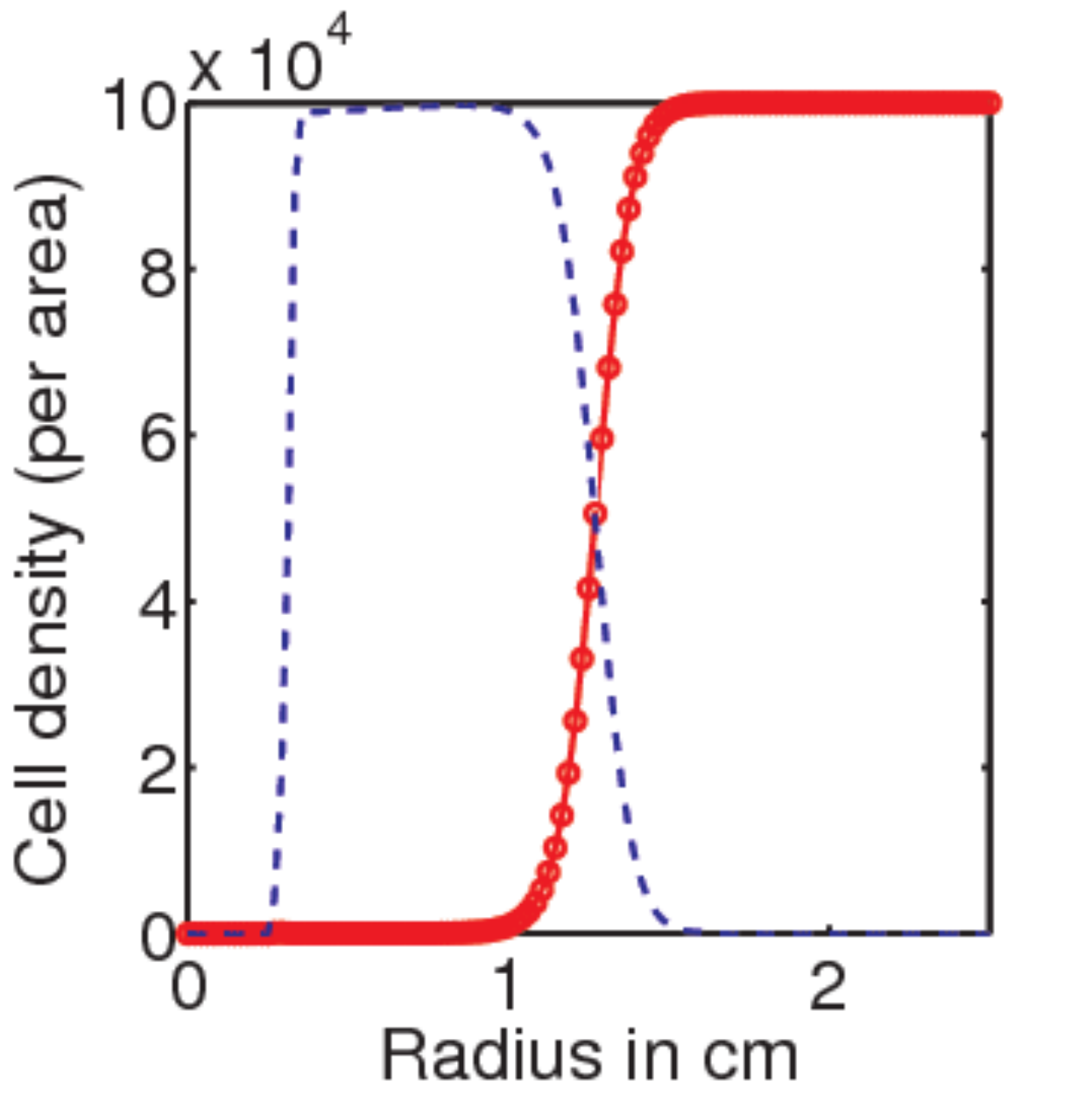}\label{den-f}} \\
   \subfigure[t=7.5 days]{\includegraphics[height=1.7in]{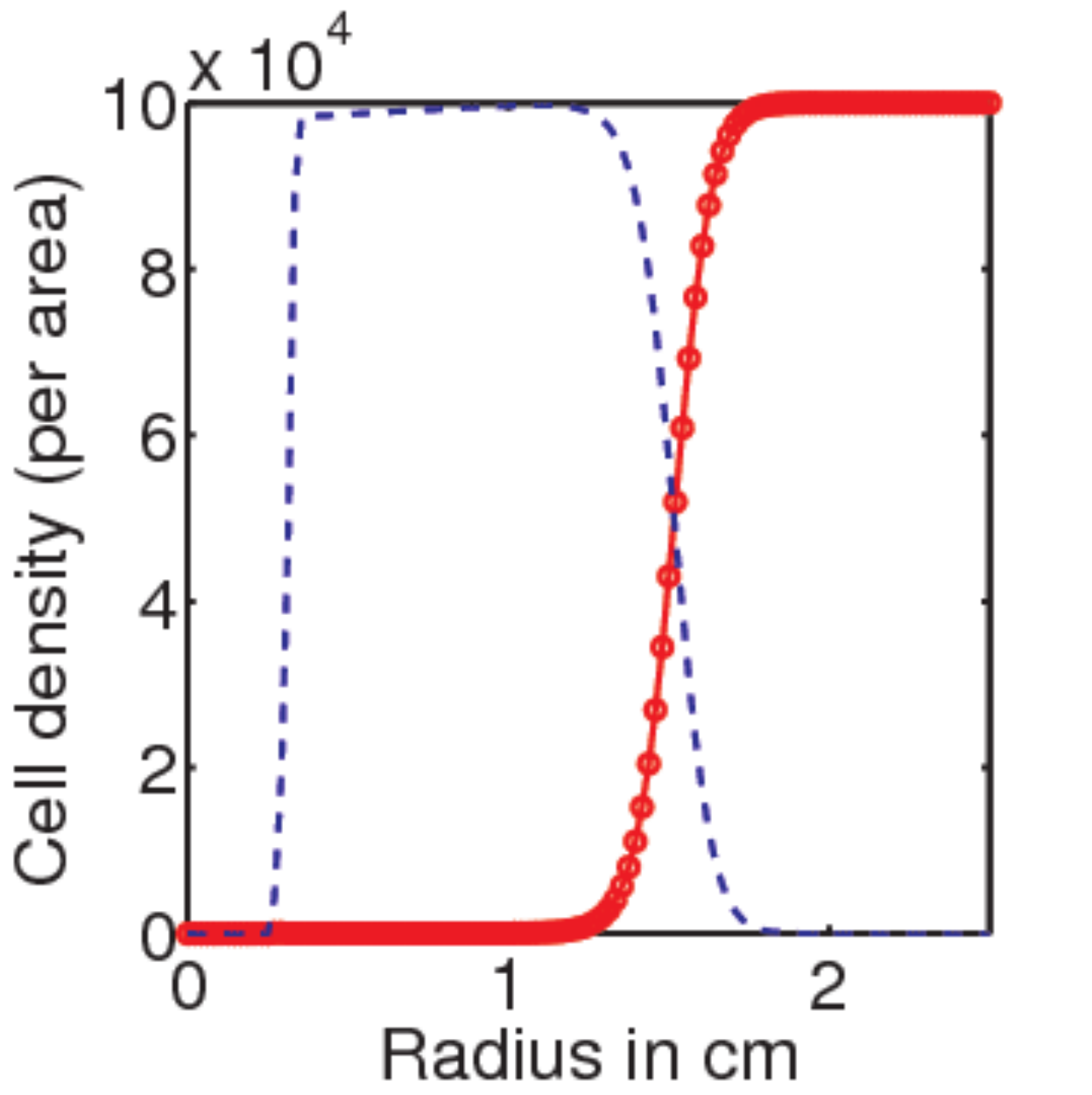}\label{den-g}}
   \subfigure[t=8.75 days]{\includegraphics[height=1.7in]{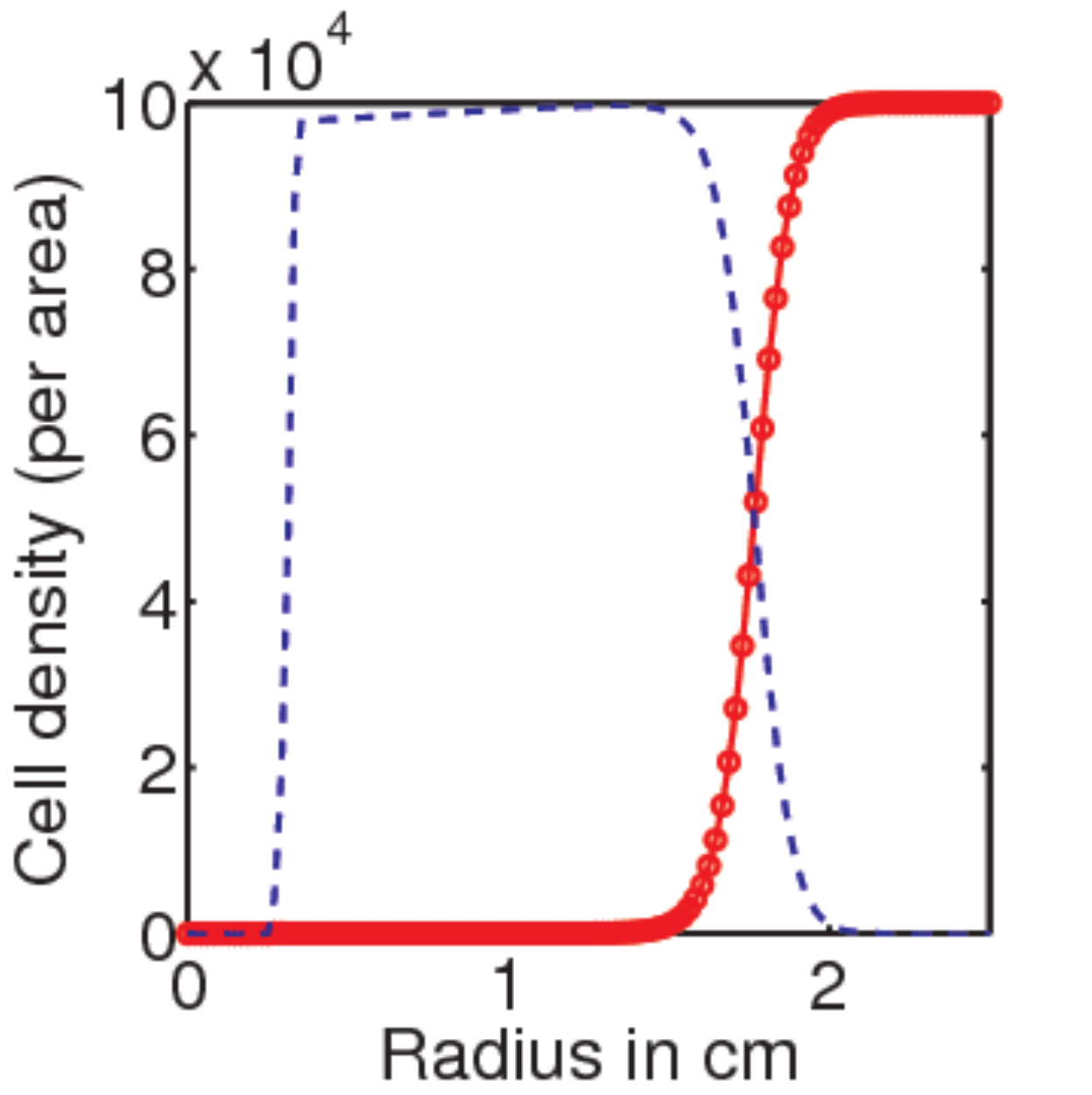}\label{den-h}}
   \subfigure[t=10]{\includegraphics[height=1.7in]{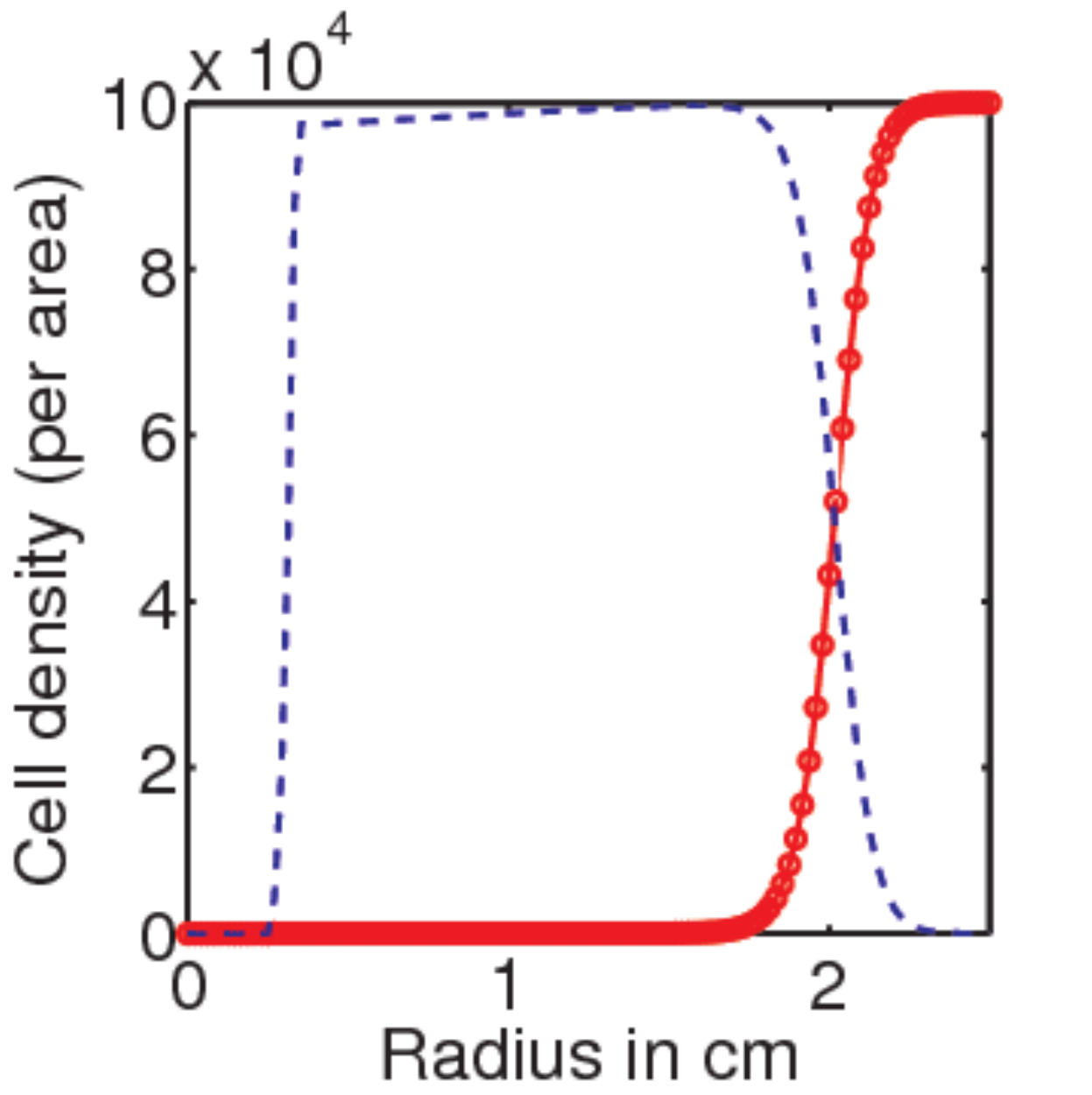}\label{den-i}}
   \caption{{\bf Healthy and Penumbra cells without EPO.} This figure shows the evolution of the healthy and penumbra (comprised of both catabolic and EPOR active) cell population densities over a period of 10 days. The results shown here correspond to a situation in which there is no production of or response to EPO. We see here that the healthy cell population is decimated. In accord with figure \ref{signaling} and the corresponding mathematical model (\ref{eq:fullSys}) this is due to the uninhibited influence of pro-inflammatory cytokines. }
   \label{noEPOdensity}
 \end{figure}

  Following the initial injury there is a quick and steady increase in the population density of catabolic and EPOR active cells and decline in the population density of healthy cells (figures \ref{den-b}-\ref{den-i}). After ten days there has formed a thick penumbra of catabolic cells which are beginning to die off (via apoptosis, see figure \ref{den-i}) showing that without the presence of EPO there is unlimited secondary pro-inflammatory cytokine-induced injury and unabated spread of the cartilage lesion.
  This corresponds qualitatively well with the expected results illustrated in figure \ref{epoTNF} (d),(e).

\begin{figure}
   \centering
   \subfigure[t=0]{\includegraphics[height=1.7in]{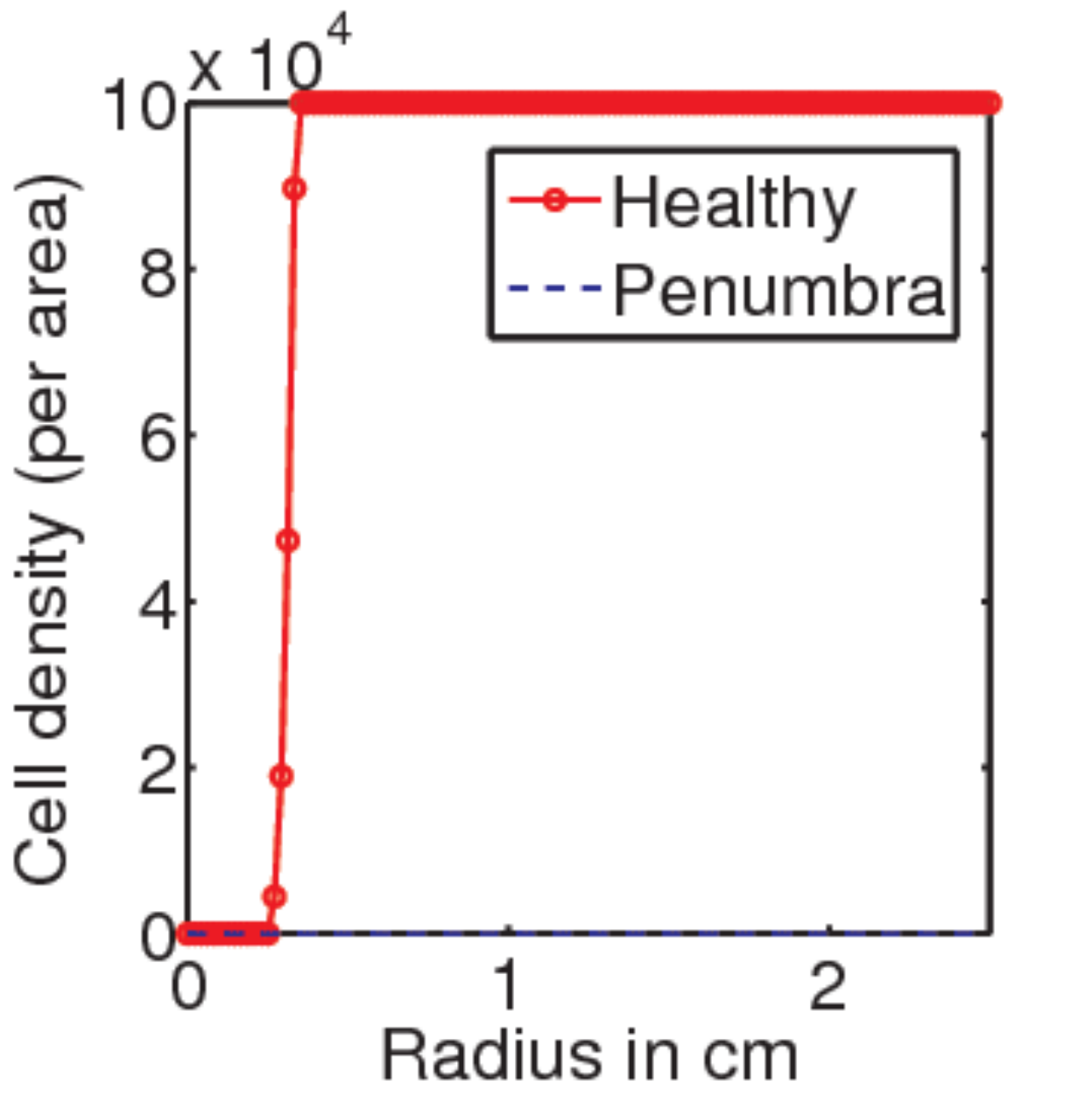}\label{de-a}}
   \subfigure[t=1.25 days]{\includegraphics[height=1.7in]{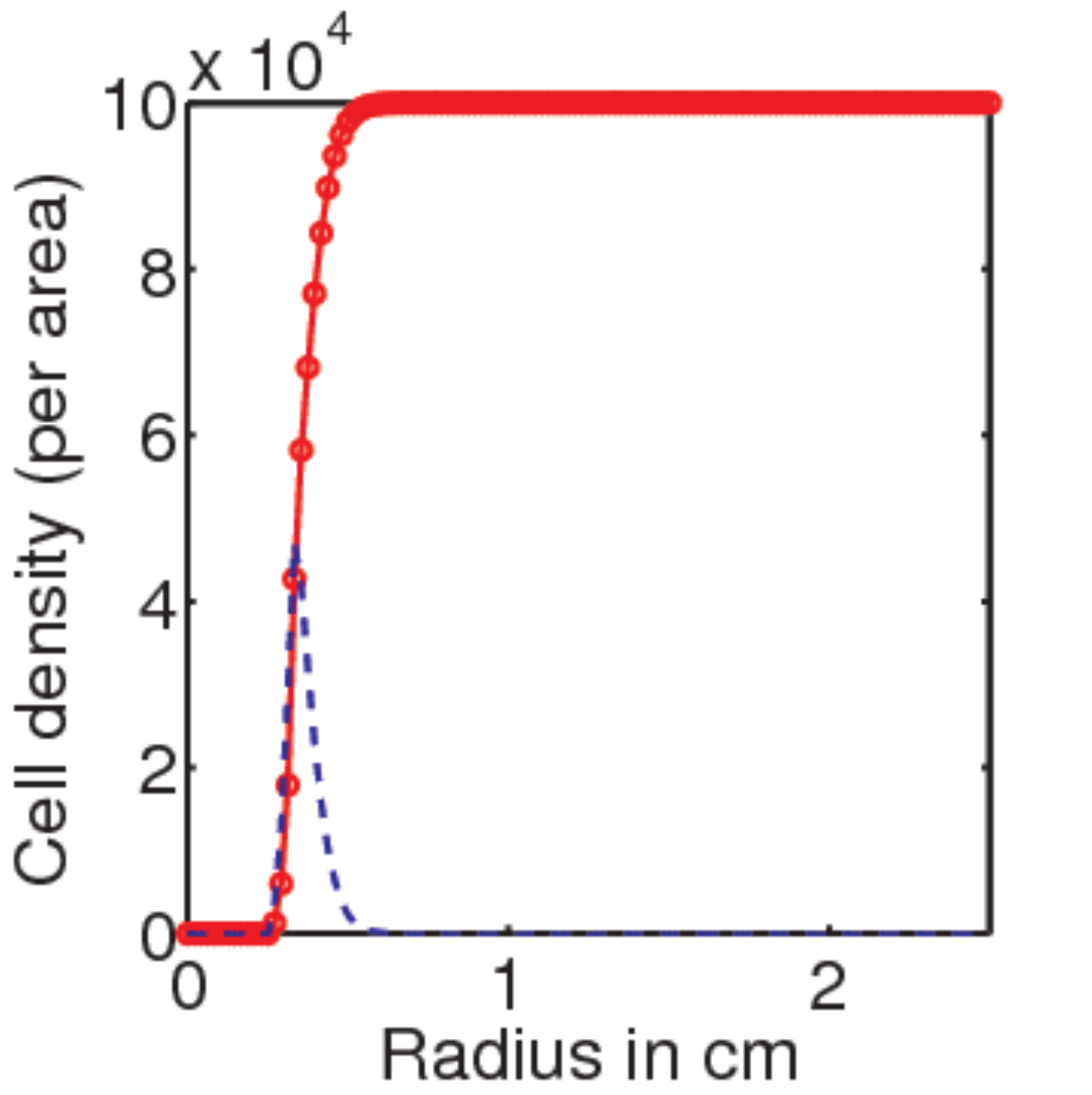}\label{de-b}}
   \subfigure[t=2.5 days]{\includegraphics[height=1.7in]{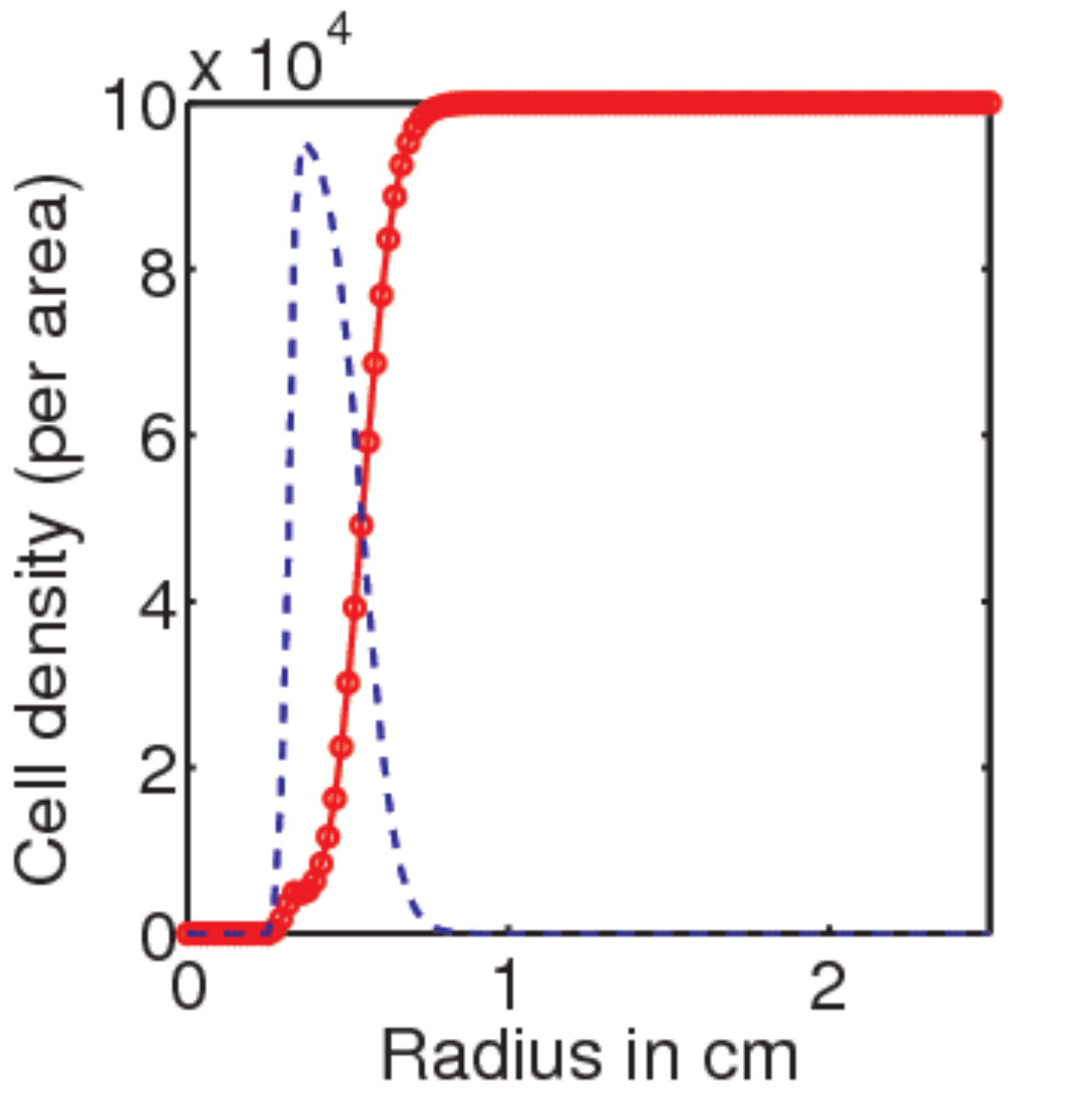}\label{de-c}}\\
   \subfigure[t=3.75 days]{\includegraphics[height=1.7in]{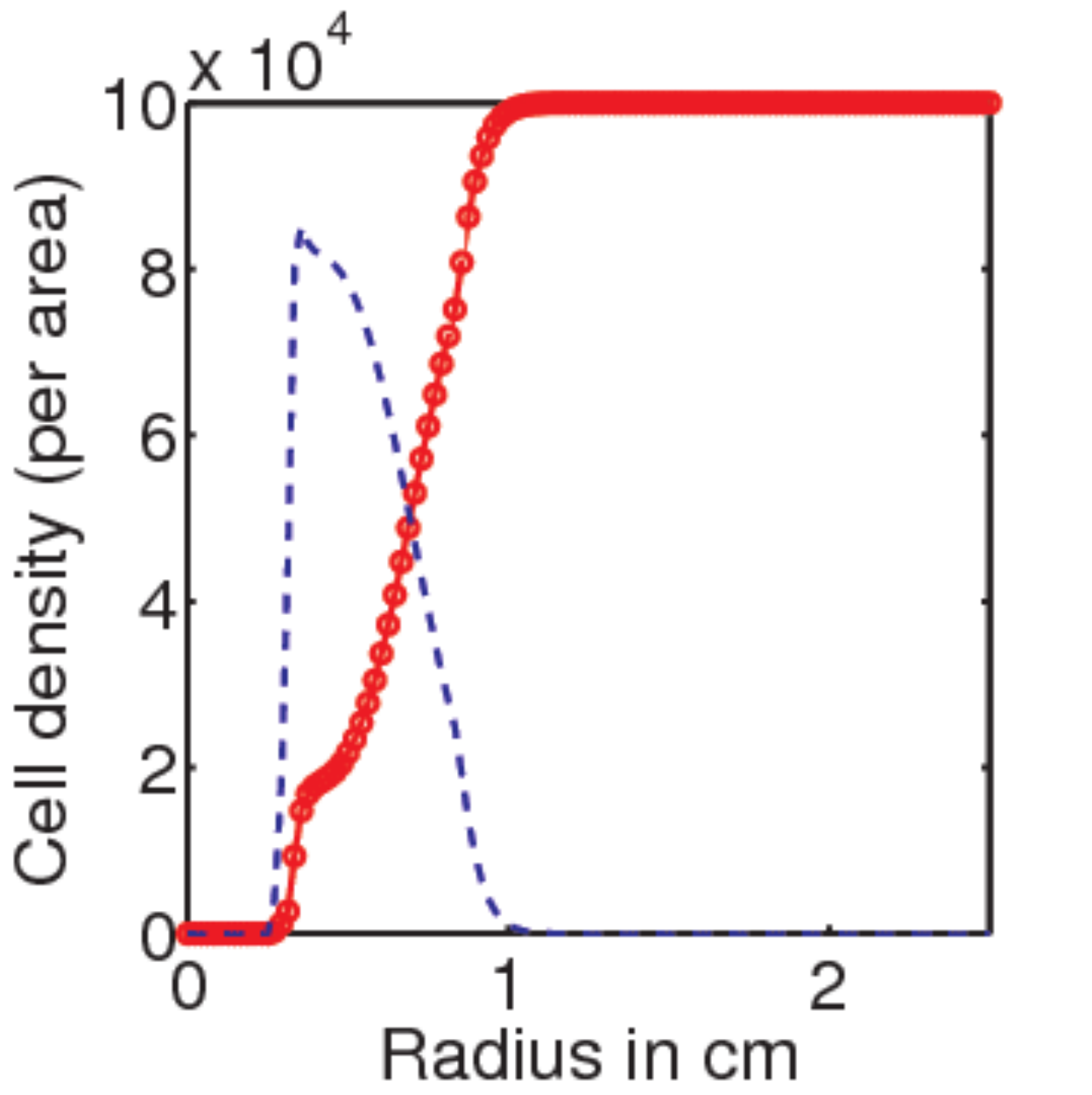}\label{de-d}}
   \subfigure[t=5 days]{\includegraphics[height=1.7in]{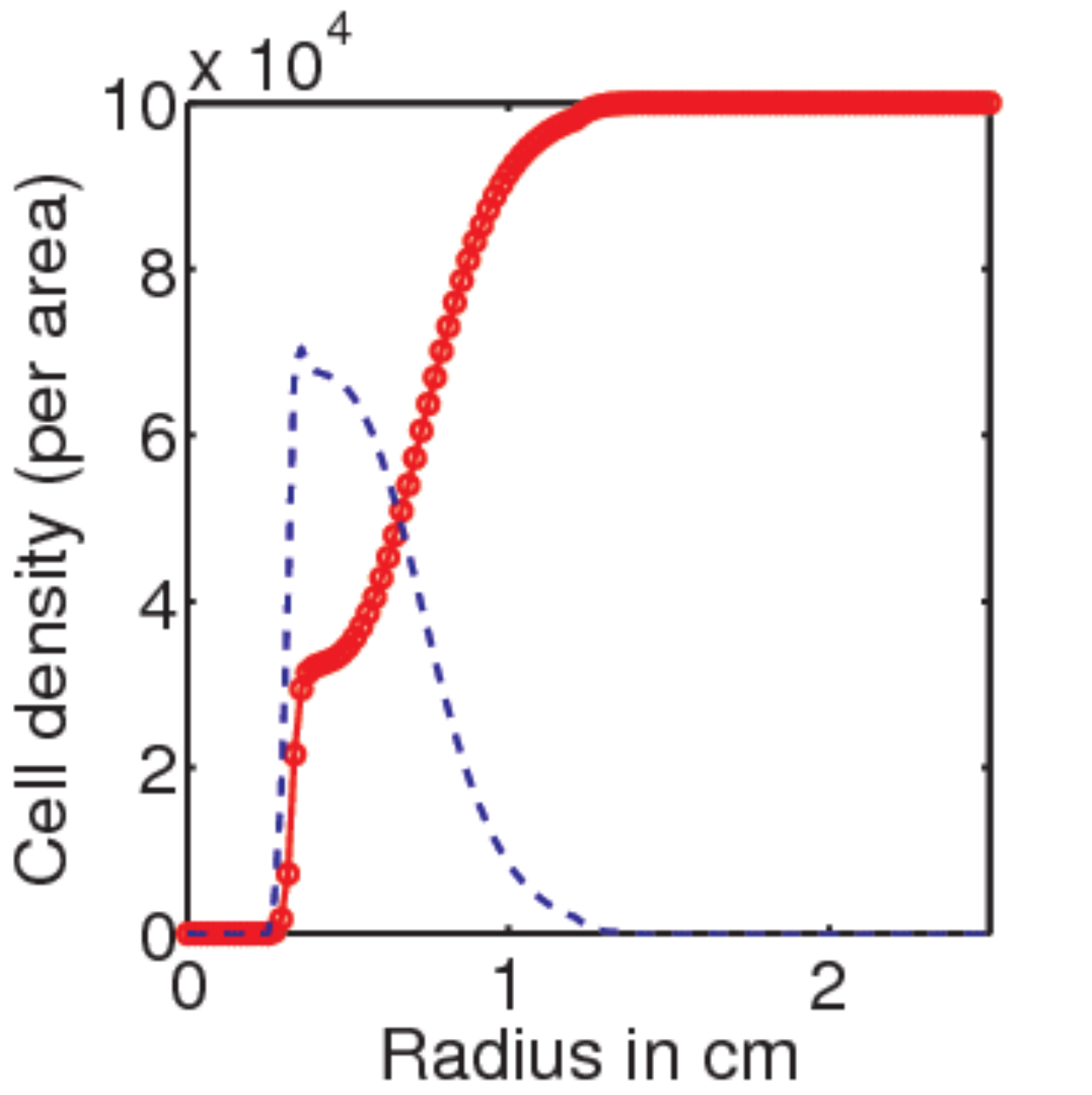}\label{de-e}}
   \subfigure[t=6.25 days]{\includegraphics[height=1.7in]{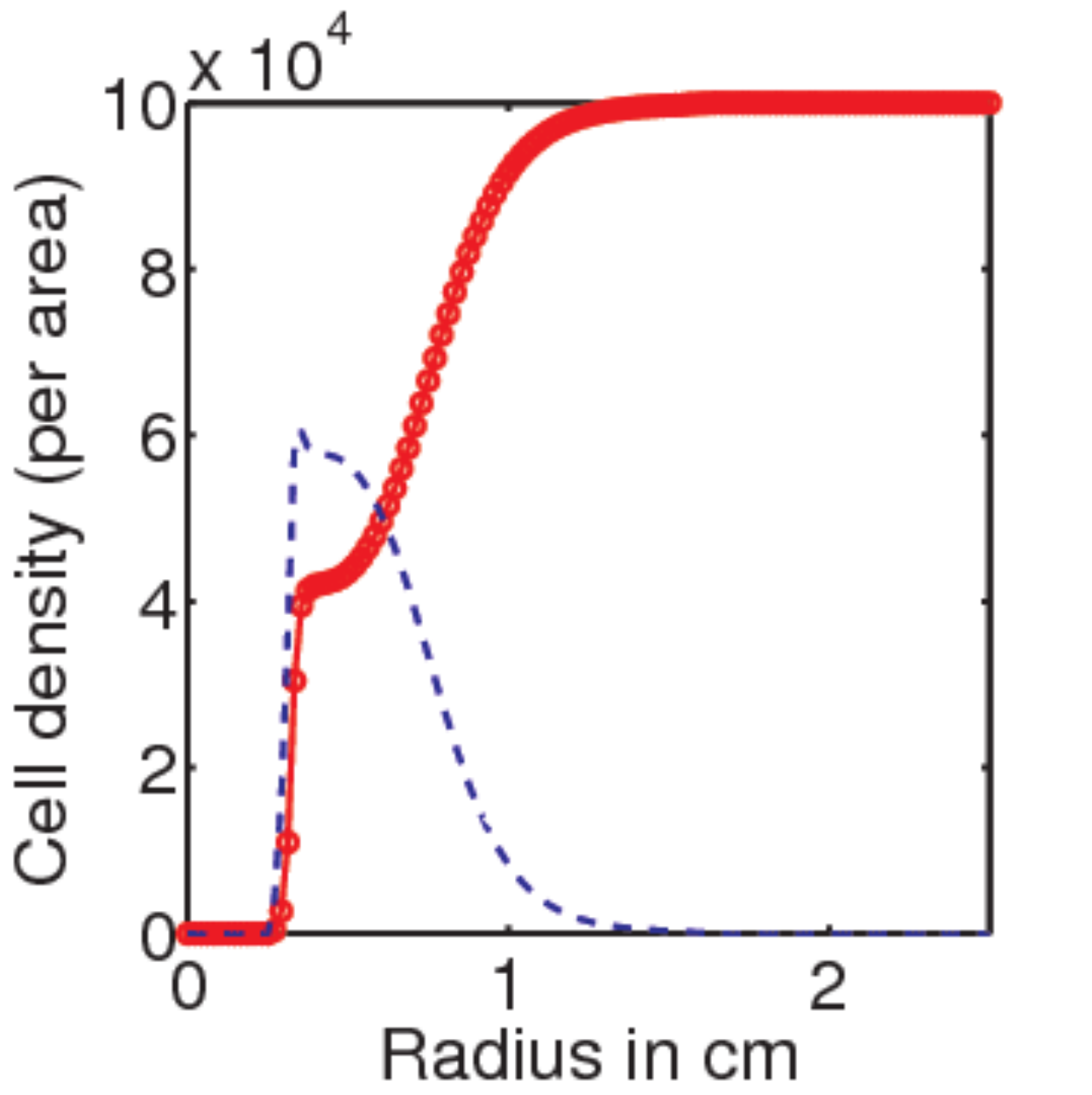}\label{de-f}} \\
   \subfigure[t=7.5 days]{\includegraphics[height=1.7in]{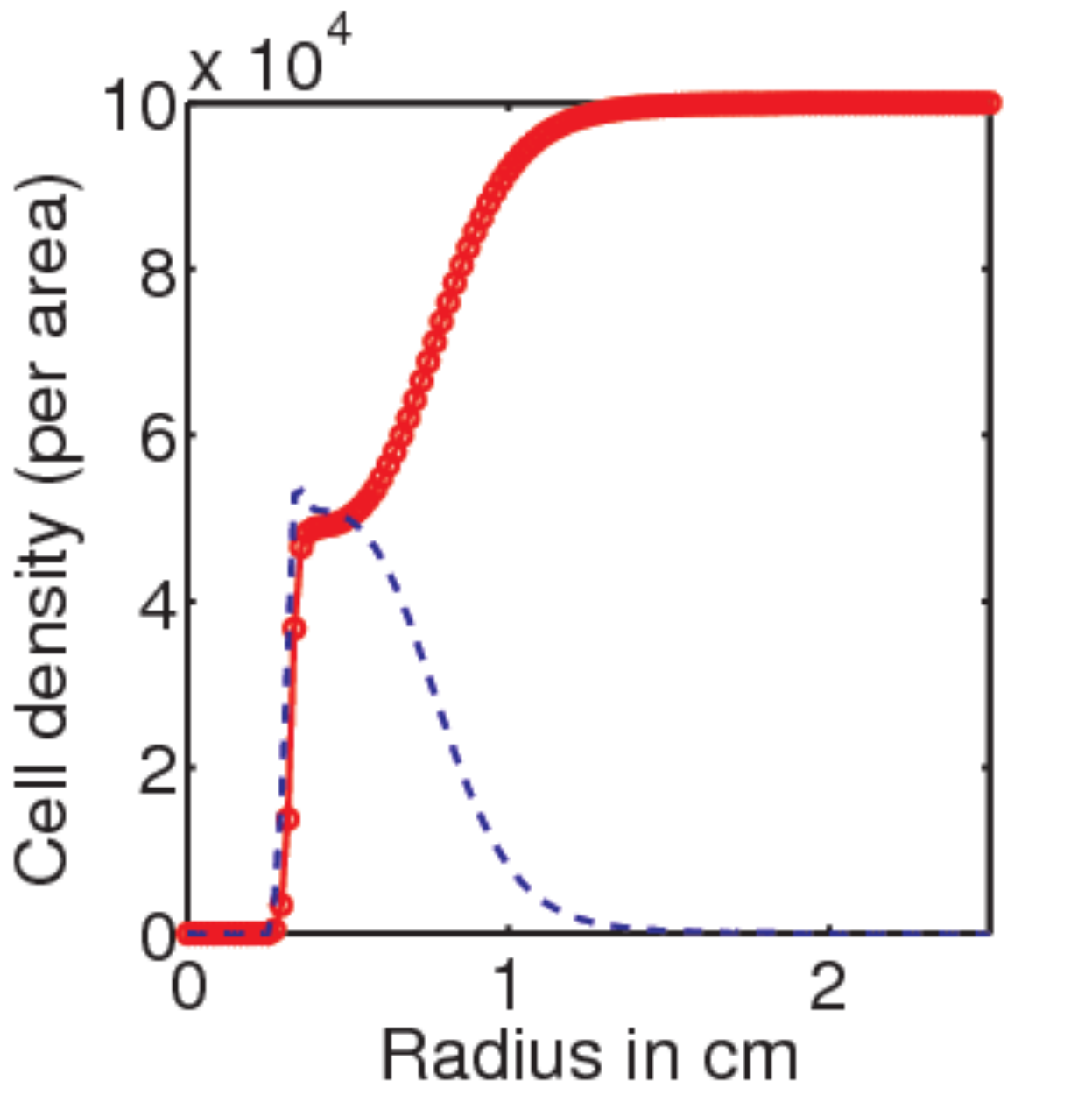}\label{de-g}}
   \subfigure[t=8.75 days]{\includegraphics[height=1.7in]{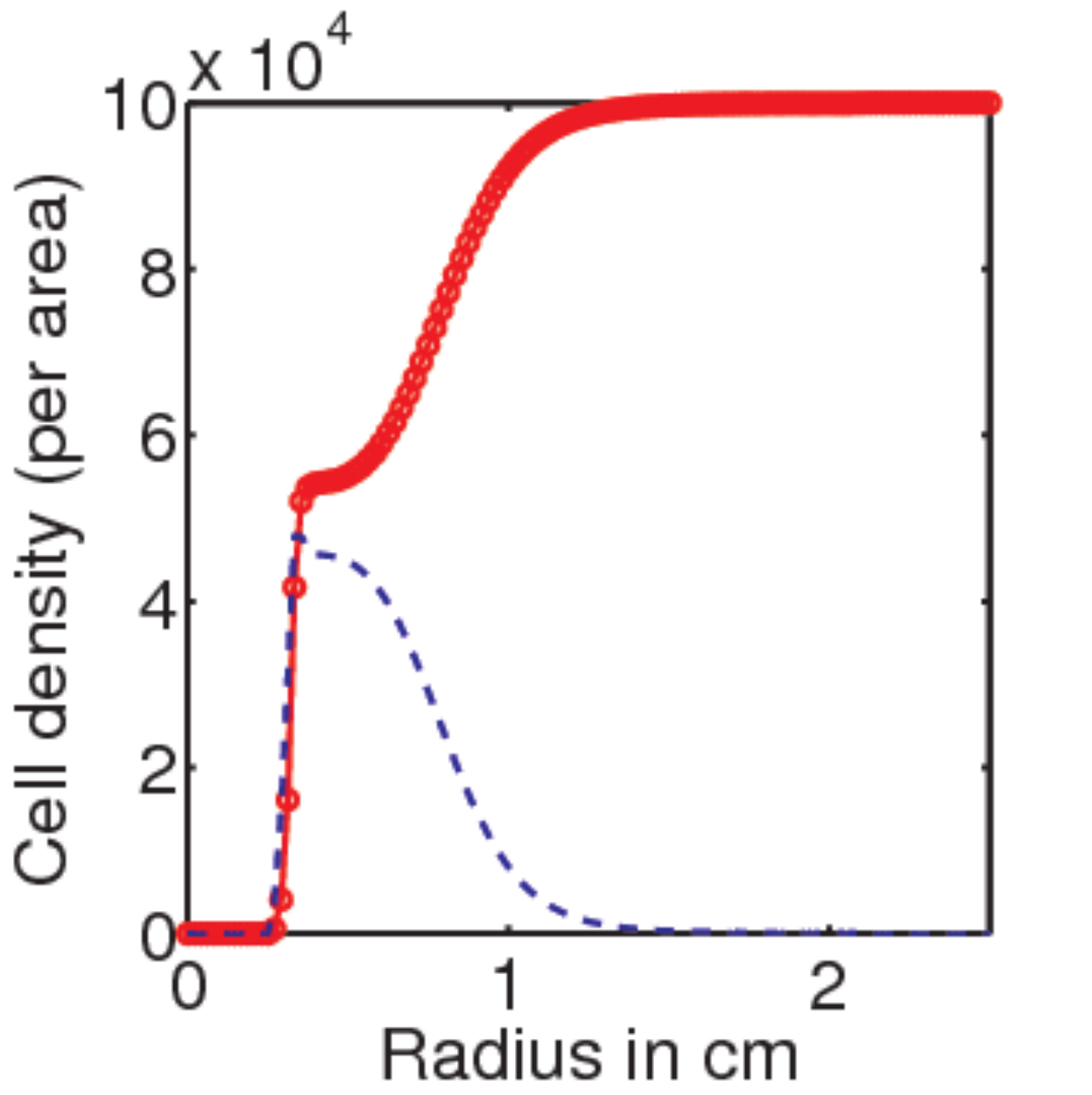}\label{de-h}}
   \subfigure[t=10]{\includegraphics[height=1.7in]{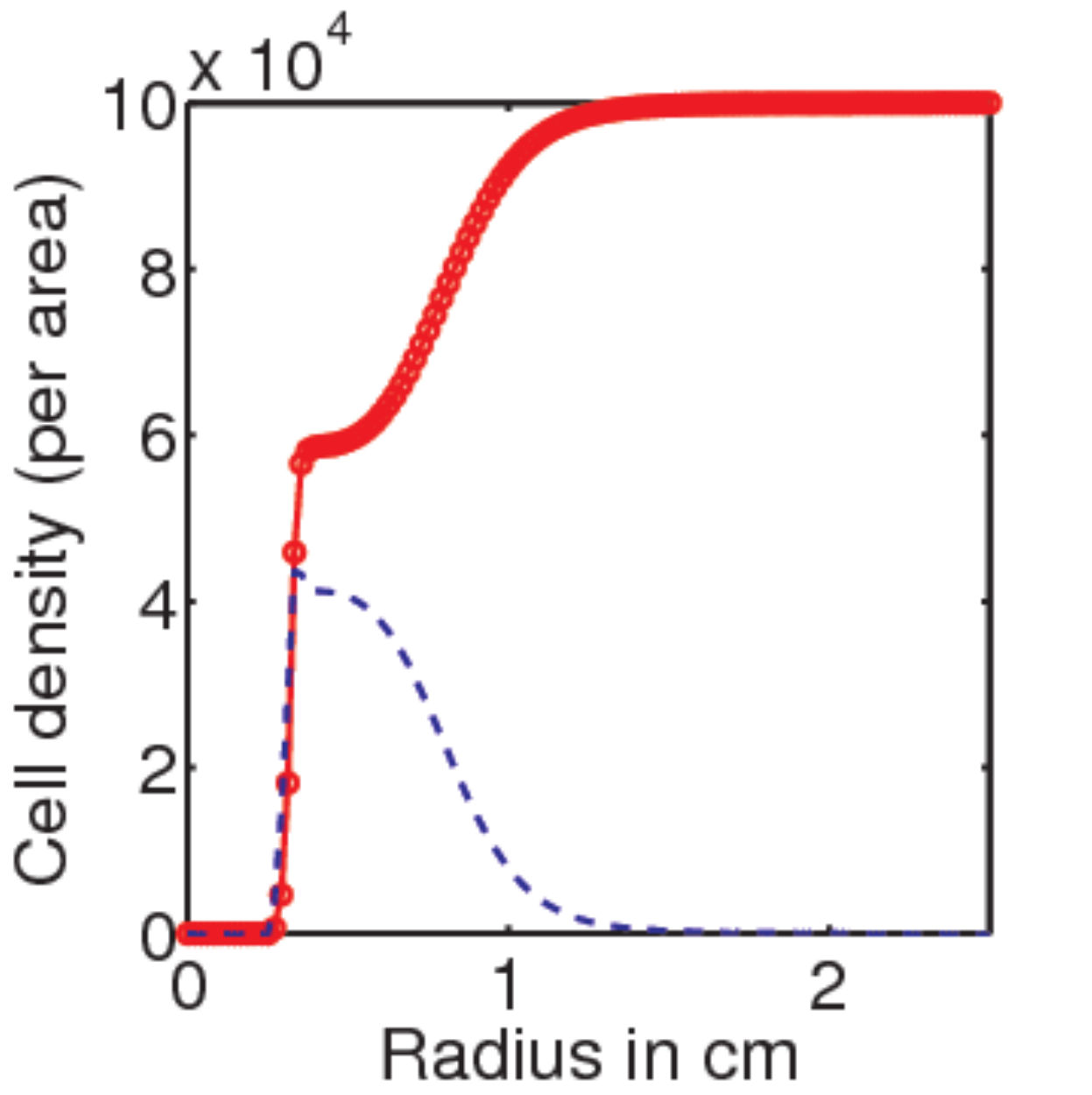}\label{de-i}}
   \caption{{\bf Healthy and Penumbra cells with EPO.} This figure shows the evolution of the healthy and penumbra cell population densities over a period of 10 days. Initially there is the development of a population of penumbra cells (\ref{de-c}) due to the time lags for EPO production and EPOR activation. However, once healthy cells at the edge of the penumbra begin to produce EPO and penumbra cells become EPOR active there is an increase in the population density of healthy cells near the injury and there ceases to be expansion of the penumbra (figures \ref{de-d} - \ref{de-i}).  }
   \label{EPOdensity}
 \end{figure}

  Figure \ref{EPOdensity} shows the time evolution of chondrocyte population density (per area, and assuming circular symmetry, as a function of radius) over a period of ten days after an initial cartilage injury with the antagonistic effects of EPO on TNF-$\alpha$. The initial conditions are the same as previously described for the simulations without EPO. The dynamics of the chondrocyte cell states population densities are the same over the first day or so as in the case with no EPO. This is due to the time delays for cells to become EPO producing and EPOR active. However, we see by day 3 or so (see figure \ref{de-d} and compare with \ref{den-d}) there is a significant difference in the population dynamics compared with that in the no EPO case. We begin to see a reversion of cells in the EPOR active state back to the healthy state. While there is still some amount of secondary injury due to the inflammatory process and presence of pro-inflammatory cytokines, there is a fixed radius (see figures \ref{de-f}-\ref{de-i}) beyond which the penumbra, and hence the lesion and secondary pro-inflammatory cytokine induced injury, cannot expand. Thus there is lesion abatement by the action of the anit-inflammatory cytokine EPO. This corresponds qualitatively well with the expected results illustrated in figure \ref{epoTNF} ($\text{d}^{\ast}$),($\text{e}^{\ast}$).

  Based on the biological assumptions summarized in the diagram shown in figure \ref{epoTNF} we have developed a mathematical representation of some of the principal features of the injury response in articular cartilage. In particular we have captured, qualitatively, secondary pro-inflammatory cytokine induced injury and lesion based on the balance between TNF-$\alpha$ and EPO as described in Brines and Cermani \cite{brines2008}. This modeling effort suggests that anti-inflammatory cytokines may act in such a way as to influence the response of chondrocytes to alarmins and pro-inflammatory cytokines. In particular sufficient concentrations of anti-inflammatory cytokines may limit healthy chondrocytes conversion to the catabolic state.  Provided dosages, dosage responses, delivery method, etc.\ the model (\ref{eq:fullSys}) can be applied to study the effectiveness of the introduction of exogenous EPO as a potential therapy for limiting secondary pro-inflammatory cytokine induced injury and promoting healing in articular cartilage.
  We note however, that due to the time delays for cells to become EPOR active, there is little or no response to EPO at early times after an injury has occurred. Thus there is a window of time during which introduction of tissue-protective EPO derivatives such as discussed in \cite{brines2008} are ineffective as therapies. This is evidenced by comparing figures \ref{den-a}-\ref{den-b} with figures \ref{de-a}-\ref{de-b} in which there is little or no difference between the case with EPO and the case without. The model presented herein also has potential applications for studying other issues related to post-traumatic stress in articular cartilage. By coupling this model with mechanical models, or including geometric features of cartilage one may be able develop a more complete theoretical description of aspects of cartilage injury of relevance in the biomedical sciences.

\section*{Acknowledgments}

This work was supported by the National Science Foundation grant number DMS-0914514 and by the National Institutes of Health Challenge
  grant number 1 RC1 AR058403-01.

\bibliography{newWoundBib}

\begin{thebibliography}{10}
\providecommand{\url}[1]{\texttt{#1}}
\providecommand{\urlprefix}{URL }
\expandafter\ifx\csname urlstyle\endcsname\relax
  \providecommand{\doi}[1]{doi:\discretionary{}{}{}#1}\else
  \providecommand{\doi}{doi:\discretionary{}{}{}\begingroup
  \urlstyle{rm}\Url}\fi
\providecommand{\bibAnnoteFile}[1]{%
  \IfFileExists{#1}{\begin{quotation}\noindent\textsc{Key:} #1\\
  \textsc{Annotation:}\ \input{#1}\end{quotation}}{}}
\providecommand{\bibAnnote}[2]{%
  \begin{quotation}\noindent\textsc{Key:} #1\\
  \textsc{Annotation:}\ #2\end{quotation}}
\providecommand{\eprint}[2][]{\url{#2}}

\bibitem{buckwalter1994}
Buckwalter J, Mow V, Ratcliffe A (1994) Restoration of injured of degenerated
  articular cartilage.
\newblock J Am Acad Orthop Surg 2: 192-201.
\bibAnnoteFile{buckwalter1994}

\bibitem{freedman2002}
Freedman K, Coleman S, Olenac C, Cole B (2002) The biology of articular
  cartilage injury and the microfracture technique for the treatment of
  articular cartilage lesions.
\newblock Sem Arthroplasty 13: 202-209.
\bibAnnoteFile{freedman2002}

\bibitem{martin2}
Goodwin W, McCabe D, Sauter E, Reese E, Walter M, et~al. (2010) Rotenone
  prevents impact-induced chondrocyte death.
\newblock J Orthop Res 28: 1057--1063.
\bibAnnoteFile{martin2}

\bibitem{martin1}
Ramakrishnan P, Hecht B, Pedersen D, Lavery M, Maynard J, et~al. (2010) Oxidant
  conditioning protects cartilage from mechanically induced damage.
\newblock J Orthop Res 28: 914--920.
\bibAnnoteFile{martin1}

\bibitem{brines2008}
Brines M, Cerami A (2008) Erythropoietin-mediated tissue protection: reducing
  collateral damage from the primary injury response.
\newblock J Intern Med 264: 405--432.
\bibAnnoteFile{brines2008}

\bibitem{harris2006}
Harris HE, Raucci A (2006) Alarming news about danger.
\newblock EMBO Reports 7: 774--778.
\bibAnnoteFile{harris2006}

\bibitem{bianchi2006}
Bianchi ME (2006) Damps, pamps and alarmins: all we need to know about danger.
\newblock J Leukoc Biol 81: 1--5.
\bibAnnoteFile{bianchi2006}

\bibitem{larry1}
Shampine LF, Reichelt MW (1997) The matlab ode suite.
\newblock SIAM J Sci Comput 18: 1--22.
\bibAnnoteFile{larry1}

\bibitem{larry2}
Shampine LF, Thompson S (2001) Solving {DDE}s in {MATLAB}.
\newblock Appl Numer Math 37: 441--458.
\bibAnnoteFile{larry2}

\bibitem{larry3}
Shampine LF, Thompson S (2009) Numerical solution of delay differential
  equations : 245--271.
\bibAnnoteFile{larry3}

\end{thebibliography}


\end{document}